\documentclass[11pt,a4paper,english,superscriptaddress,aps,preprint]{revtex4}
\usepackage{amsmath}
\usepackage{amssymb}
\usepackage{graphicx}
\usepackage{slashed}
\makeatletter
\usepackage{babel}
\newcommand{\bea}{\begin{eqnarray}}
\newcommand{\eea}{\end{eqnarray}}

\newcommand{\pa}{\partial}
\renewcommand{\a}{\alpha}

\renewcommand{\b}{\beta}

\newcommand{\q}{\theta}

\newcommand{\be}{\begin{equation}}
\newcommand{\ee}{\end{equation}}
\begin{document}
\immediate\write16{<<WARNING: LINEDRAW macros work with emTeX-dvivers
                    and other drivers supporting emTeX \special's
                    (dviscr, dvihplj, dvidot, dvips, dviwin, etc.) >>}

%% Macros for drawing Feynman graphs and other complex diagrams
%% Designed by A.V.Voronin (1993); modified in 1995
%% Steklov Math. Inst., e-mail: av@voronin.mian.su
%%
\newdimen\Lengthunit       \Lengthunit  = 1.5cm
\newcount\Nhalfperiods     \Nhalfperiods= 9
\newcount\magnitude        \magnitude = 1000

\catcode`\*=11
\newdimen\L*   \newdimen\d*   \newdimen\d**
\newdimen\dm*  \newdimen\dd*  \newdimen\dt*
\newdimen\a*   \newdimen\b*   \newdimen\c*
\newdimen\a**  \newdimen\b**
\newdimen\xL*  \newdimen\yL*
\newdimen\rx*  \newdimen\ry*
\newdimen\tmp* \newdimen\linwid*

\newcount\k*   \newcount\l*   \newcount\m*
\newcount\k**  \newcount\l**  \newcount\m**
\newcount\n*   \newcount\dn*  \newcount\r*
\newcount\N*   \newcount\*one \newcount\*two  \*one=1 \*two=2
\newcount\*ths \*ths=1000
\newcount\angle*  \newcount\q*  \newcount\q**
\newcount\angle** \angle**=0
\newcount\sc*     \sc*=0

\newtoks\cos*  \cos*={1}
\newtoks\sin*  \sin*={0}

\catcode`\[=13

\def\rotate(#1){\advance\angle**#1\angle*=\angle**
\q**=\angle*\ifnum\q**<0\q**=-\q**\fi
\ifnum\q**>360\q*=\angle*\divide\q*360\multiply\q*360\advance\angle*-\q*\fi
\ifnum\angle*<0\advance\angle*360\fi\q**=\angle*\divide\q**90\q**=\q**
\def\sgcos*{+}\def\sgsin*{+}\relax
\ifcase\q**\or
 \def\sgcos*{-}\def\sgsin*{+}\or
 \def\sgcos*{-}\def\sgsin*{-}\or
 \def\sgcos*{+}\def\sgsin*{-}\else\fi
\q*=\q**
\multiply\q*90\advance\angle*-\q*
\ifnum\angle*>45\sc*=1\angle*=-\angle*\advance\angle*90\else\sc*=0\fi
\def[##1,##2]{\ifnum\sc*=0\relax
\edef\cs*{\sgcos*.##1}\edef\sn*{\sgsin*.##2}\ifcase\q**\or
 \edef\cs*{\sgcos*.##2}\edef\sn*{\sgsin*.##1}\or
 \edef\cs*{\sgcos*.##1}\edef\sn*{\sgsin*.##2}\or
 \edef\cs*{\sgcos*.##2}\edef\sn*{\sgsin*.##1}\else\fi\else
\edef\cs*{\sgcos*.##2}\edef\sn*{\sgsin*.##1}\ifcase\q**\or
 \edef\cs*{\sgcos*.##1}\edef\sn*{\sgsin*.##2}\or
 \edef\cs*{\sgcos*.##2}\edef\sn*{\sgsin*.##1}\or
 \edef\cs*{\sgcos*.##1}\edef\sn*{\sgsin*.##2}\else\fi\fi
\cos*={\cs*}\sin*={\sn*}\global\edef\gcos*{\cs*}\global\edef\gsin*{\sn*}}\relax
\ifcase\angle*[9999,0]\or
[999,017]\or[999,034]\or[998,052]\or[997,069]\or[996,087]\or
[994,104]\or[992,121]\or[990,139]\or[987,156]\or[984,173]\or
[981,190]\or[978,207]\or[974,224]\or[970,241]\or[965,258]\or
[961,275]\or[956,292]\or[951,309]\or[945,325]\or[939,342]\or
[933,358]\or[927,374]\or[920,390]\or[913,406]\or[906,422]\or
[898,438]\or[891,453]\or[882,469]\or[874,484]\or[866,499]\or
[857,515]\or[848,529]\or[838,544]\or[829,559]\or[819,573]\or
[809,587]\or[798,601]\or[788,615]\or[777,629]\or[766,642]\or
[754,656]\or[743,669]\or[731,681]\or[719,694]\or[707,707]\or
\else[9999,0]\fi}

\catcode`\[=12

\def\GRAPH(hsize=#1)#2{\hbox to #1\Lengthunit{#2\hss}}

\def\Linewidth#1{\global\linwid*=#1\relax
\global\divide\linwid*10\global\multiply\linwid*\mag
\global\divide\linwid*100\special{em:linewidth \the\linwid*}}

\Linewidth{.4pt}
\def\sm*{\special{em:moveto}}
\def\sl*{\special{em:lineto}}
\let\moveto=\sm*
\let\lineto=\sl*
\newbox\spm*   \newbox\spl*
\setbox\spm*\hbox{\sm*}
\setbox\spl*\hbox{\sl*}

\def\mov#1(#2,#3)#4{\rlap{\L*=#1\Lengthunit
\xL*=#2\L* \yL*=#3\L*
\xL*=\xscale\xL* \yL*=\yscale\yL*
\rx* \the\cos*\xL* \tmp* \the\sin*\yL* \advance\rx*-\tmp*
\ry* \the\cos*\yL* \tmp* \the\sin*\xL* \advance\ry*\tmp*
\kern\rx*\raise\ry*\hbox{#4}}}

\def\rmov*(#1,#2)#3{\rlap{\xL*=#1\yL*=#2\relax
\rx* \the\cos*\xL* \tmp* \the\sin*\yL* \advance\rx*-\tmp*
\ry* \the\cos*\yL* \tmp* \the\sin*\xL* \advance\ry*\tmp*
\kern\rx*\raise\ry*\hbox{#3}}}

\def\lin#1(#2,#3){\rlap{\sm*\mov#1(#2,#3){\sl*}}}

\def\arr*(#1,#2,#3){\rmov*(#1\dd*,#1\dt*){\sm*
\rmov*(#2\dd*,#2\dt*){\rmov*(#3\dt*,-#3\dd*){\sl*}}\sm*
\rmov*(#2\dd*,#2\dt*){\rmov*(-#3\dt*,#3\dd*){\sl*}}}}

\def\arrow#1(#2,#3){\rlap{\lin#1(#2,#3)\mov#1(#2,#3){\relax
\d**=-.012\Lengthunit\dd*=#2\d**\dt*=#3\d**
\arr*(1,10,4)\arr*(3,8,4)\arr*(4.8,4.2,3)}}}

\def\arrlin#1(#2,#3){\rlap{\L*=#1\Lengthunit\L*=.5\L*
\lin#1(#2,#3)\rmov*(#2\L*,#3\L*){\arrow.1(#2,#3)}}}

\def\dasharrow#1(#2,#3){\rlap{{\Lengthunit=0.9\Lengthunit
\dashlin#1(#2,#3)\mov#1(#2,#3){\sm*}}\mov#1(#2,#3){\sl*
\d**=-.012\Lengthunit\dd*=#2\d**\dt*=#3\d**
\arr*(1,10,4)\arr*(3,8,4)\arr*(4.8,4.2,3)}}}

\def\clap#1{\hbox to 0pt{\hss #1\hss}}

\def\ind(#1,#2)#3{\rlap{\L*=.1\Lengthunit
\xL*=#1\L* \yL*=#2\L*
\rx* \the\cos*\xL* \tmp* \the\sin*\yL* \advance\rx*-\tmp*
\ry* \the\cos*\yL* \tmp* \the\sin*\xL* \advance\ry*\tmp*
\kern\rx*\raise\ry*\hbox{\lower2pt\clap{$#3$}}}}

\def\sh*(#1,#2)#3{\rlap{\dm*=\the\n*\d**
\xL*=\xscale\dm* \yL*=\yscale\dm* \xL*=#1\xL* \yL*=#2\yL*
\rx* \the\cos*\xL* \tmp* \the\sin*\yL* \advance\rx*-\tmp*
\ry* \the\cos*\yL* \tmp* \the\sin*\xL* \advance\ry*\tmp*
\kern\rx*\raise\ry*\hbox{#3}}}

\def\calcnum*#1(#2,#3){\a*=1000sp\b*=1000sp\a*=#2\a*\b*=#3\b*
\ifdim\a*<0pt\a*-\a*\fi\ifdim\b*<0pt\b*-\b*\fi
\ifdim\a*>\b*\c*=.96\a*\advance\c*.4\b*
\else\c*=.96\b*\advance\c*.4\a*\fi
\k*\a*\multiply\k*\k*\l*\b*\multiply\l*\l*
\m*\k*\advance\m*\l*\n*\c*\r*\n*\multiply\n*\n*
\dn*\m*\advance\dn*-\n*\divide\dn*2\divide\dn*\r*
\advance\r*\dn*
\c*=\the\Nhalfperiods5sp\c*=#1\c*\ifdim\c*<0pt\c*-\c*\fi
\multiply\c*\r*\N*\c*\divide\N*10000}

\def\dashlin#1(#2,#3){\rlap{\calcnum*#1(#2,#3)\relax
\d**=#1\Lengthunit\ifdim\d**<0pt\d**-\d**\fi
\divide\N*2\multiply\N*2\advance\N*\*one
\divide\d**\N*\sm*\n*\*one\sh*(#2,#3){\sl*}\loop
\advance\n*\*one\sh*(#2,#3){\sm*}\advance\n*\*one
\sh*(#2,#3){\sl*}\ifnum\n*<\N*\repeat}}

\def\dashdotlin#1(#2,#3){\rlap{\calcnum*#1(#2,#3)\relax
\d**=#1\Lengthunit\ifdim\d**<0pt\d**-\d**\fi
\divide\N*2\multiply\N*2\advance\N*1\multiply\N*2\relax
\divide\d**\N*\sm*\n*\*two\sh*(#2,#3){\sl*}\loop
\advance\n*\*one\sh*(#2,#3){\kern-1.48pt\lower.5pt\hbox{\rm.}}\relax
\advance\n*\*one\sh*(#2,#3){\sm*}\advance\n*\*two
\sh*(#2,#3){\sl*}\ifnum\n*<\N*\repeat}}

\def\shl*(#1,#2)#3{\kern#1#3\lower#2#3\hbox{\unhcopy\spl*}}

\def\trianglin#1(#2,#3){\rlap{\toks0={#2}\toks1={#3}\calcnum*#1(#2,#3)\relax
\dd*=.57\Lengthunit\dd*=#1\dd*\divide\dd*\N*
\divide\dd*\*ths \multiply\dd*\magnitude
\d**=#1\Lengthunit\ifdim\d**<0pt\d**-\d**\fi
\multiply\N*2\divide\d**\N*\sm*\n*\*one\loop
\shl**{\dd*}\dd*-\dd*\advance\n*2\relax
\ifnum\n*<\N*\repeat\n*\N*\shl**{0pt}}}

\def\wavelin#1(#2,#3){\rlap{\toks0={#2}\toks1={#3}\calcnum*#1(#2,#3)\relax
\dd*=.23\Lengthunit\dd*=#1\dd*\divide\dd*\N*
\divide\dd*\*ths \multiply\dd*\magnitude
\d**=#1\Lengthunit\ifdim\d**<0pt\d**-\d**\fi
\multiply\N*4\divide\d**\N*\sm*\n*\*one\loop
\shl**{\dd*}\dt*=1.3\dd*\advance\n*\*one
\shl**{\dt*}\advance\n*\*one
\shl**{\dd*}\advance\n*\*two
\dd*-\dd*\ifnum\n*<\N*\repeat\n*\N*\shl**{0pt}}}

\def\w*lin(#1,#2){\rlap{\toks0={#1}\toks1={#2}\d**=\Lengthunit\dd*=-.12\d**
\divide\dd*\*ths \multiply\dd*\magnitude
\N*8\divide\d**\N*\sm*\n*\*one\loop
\shl**{\dd*}\dt*=1.3\dd*\advance\n*\*one
\shl**{\dt*}\advance\n*\*one
\shl**{\dd*}\advance\n*\*one
\shl**{0pt}\dd*-\dd*\advance\n*1\ifnum\n*<\N*\repeat}}

\def\l*arc(#1,#2)[#3][#4]{\rlap{\toks0={#1}\toks1={#2}\d**=\Lengthunit
\dd*=#3.037\d**\dd*=#4\dd*\dt*=#3.049\d**\dt*=#4\dt*\ifdim\d**>10mm\relax
\d**=.25\d**\n*\*one\shl**{-\dd*}\n*\*two\shl**{-\dt*}\n*3\relax
\shl**{-\dd*}\n*4\relax\shl**{0pt}\else
\ifdim\d**>5mm\d**=.5\d**\n*\*one\shl**{-\dt*}\n*\*two
\shl**{0pt}\else\n*\*one\shl**{0pt}\fi\fi}}

\def\d*arc(#1,#2)[#3][#4]{\rlap{\toks0={#1}\toks1={#2}\d**=\Lengthunit
\dd*=#3.037\d**\dd*=#4\dd*\d**=.25\d**\sm*\n*\*one\shl**{-\dd*}\relax
\n*3\relax\sh*(#1,#2){\xL*=\xscale\dd*\yL*=\yscale\dd*
\kern#2\xL*\lower#1\yL*\hbox{\sm*}}\n*4\relax\shl**{0pt}}}

\def\shl**#1{\c*=\the\n*\d**\d*=#1\relax
\a*=\the\toks0\c*\b*=\the\toks1\d*\advance\a*-\b*
\b*=\the\toks1\c*\d*=\the\toks0\d*\advance\b*\d*
\a*=\xscale\a*\b*=\yscale\b*
\rx* \the\cos*\a* \tmp* \the\sin*\b* \advance\rx*-\tmp*
\ry* \the\cos*\b* \tmp* \the\sin*\a* \advance\ry*\tmp*
\raise\ry*\rlap{\kern\rx*\unhcopy\spl*}}

\def\wlin*#1(#2,#3)[#4]{\rlap{\toks0={#2}\toks1={#3}\relax
\c*=#1\l*\c*\c*=.01\Lengthunit\m*\c*\divide\l*\m*
\c*=\the\Nhalfperiods5sp\multiply\c*\l*\N*\c*\divide\N*\*ths
\divide\N*2\multiply\N*2\advance\N*\*one
\dd*=.002\Lengthunit\dd*=#4\dd*\multiply\dd*\l*\divide\dd*\N*
\divide\dd*\*ths \multiply\dd*\magnitude
\d**=#1\multiply\N*4\divide\d**\N*\sm*\n*\*one\loop
\shl**{\dd*}\dt*=1.3\dd*\advance\n*\*one
\shl**{\dt*}\advance\n*\*one
\shl**{\dd*}\advance\n*\*two
\dd*-\dd*\ifnum\n*<\N*\repeat\n*\N*\shl**{0pt}}}

\def\wavebox#1{\setbox0\hbox{#1}\relax
\a*=\wd0\advance\a*14pt\b*=\ht0\advance\b*\dp0\advance\b*14pt\relax
\hbox{\kern9pt\relax
\rmov*(0pt,\ht0){\rmov*(-7pt,7pt){\wlin*\a*(1,0)[+]\wlin*\b*(0,-1)[-]}}\relax
\rmov*(\wd0,-\dp0){\rmov*(7pt,-7pt){\wlin*\a*(-1,0)[+]\wlin*\b*(0,1)[-]}}\relax
\box0\kern9pt}}

\def\rectangle#1(#2,#3){\relax
\lin#1(#2,0)\lin#1(0,#3)\mov#1(0,#3){\lin#1(#2,0)}\mov#1(#2,0){\lin#1(0,#3)}}

\def\dashrectangle#1(#2,#3){\dashlin#1(#2,0)\dashlin#1(0,#3)\relax
\mov#1(0,#3){\dashlin#1(#2,0)}\mov#1(#2,0){\dashlin#1(0,#3)}}

\def\waverectangle#1(#2,#3){\L*=#1\Lengthunit\a*=#2\L*\b*=#3\L*
\ifdim\a*<0pt\a*-\a*\def\x*{-1}\else\def\x*{1}\fi
\ifdim\b*<0pt\b*-\b*\def\y*{-1}\else\def\y*{1}\fi
\wlin*\a*(\x*,0)[-]\wlin*\b*(0,\y*)[+]\relax
\mov#1(0,#3){\wlin*\a*(\x*,0)[+]}\mov#1(#2,0){\wlin*\b*(0,\y*)[-]}}

\def\calcparab*{\ifnum\n*>\m*\k*\N*\advance\k*-\n*\else\k*\n*\fi
\a*=\the\k* sp\a*=10\a*\b*\dm*\advance\b*-\a*\k*\b*
\a*=\the\*ths\b*\divide\a*\l*\multiply\a*\k*
\divide\a*\l*\k*\*ths\r*\a*\advance\k*-\r*\dt*=\the\k*\L*}

\def\arcto#1(#2,#3)[#4]{\rlap{\toks0={#2}\toks1={#3}\calcnum*#1(#2,#3)\relax
\dm*=135sp\dm*=#1\dm*\d**=#1\Lengthunit\ifdim\dm*<0pt\dm*-\dm*\fi
\multiply\dm*\r*\a*=.3\dm*\a*=#4\a*\ifdim\a*<0pt\a*-\a*\fi
\advance\dm*\a*\N*\dm*\divide\N*10000\relax
\divide\N*2\multiply\N*2\advance\N*\*one
\L*=-.25\d**\L*=#4\L*\divide\d**\N*\divide\L*\*ths
\m*\N*\divide\m*2\dm*=\the\m*5sp\l*\dm*\sm*\n*\*one\loop
\calcparab*\shl**{-\dt*}\advance\n*1\ifnum\n*<\N*\repeat}}

\def\arrarcto#1(#2,#3)[#4]{\L*=#1\Lengthunit\L*=.54\L*
\arcto#1(#2,#3)[#4]\rmov*(#2\L*,#3\L*){\d*=.457\L*\d*=#4\d*\d**-\d*
\rmov*(#3\d**,#2\d*){\arrow.02(#2,#3)}}}

\def\dasharcto#1(#2,#3)[#4]{\rlap{\toks0={#2}\toks1={#3}\relax
\calcnum*#1(#2,#3)\dm*=\the\N*5sp\a*=.3\dm*\a*=#4\a*\ifdim\a*<0pt\a*-\a*\fi
\advance\dm*\a*\N*\dm*
\divide\N*20\multiply\N*2\advance\N*1\d**=#1\Lengthunit
\L*=-.25\d**\L*=#4\L*\divide\d**\N*\divide\L*\*ths
\m*\N*\divide\m*2\dm*=\the\m*5sp\l*\dm*
\sm*\n*\*one\loop\calcparab*
\shl**{-\dt*}\advance\n*1\ifnum\n*>\N*\else\calcparab*
\sh*(#2,#3){\xL*=#3\dt* \yL*=#2\dt*
\rx* \the\cos*\xL* \tmp* \the\sin*\yL* \advance\rx*\tmp*
\ry* \the\cos*\yL* \tmp* \the\sin*\xL* \advance\ry*-\tmp*
\kern\rx*\lower\ry*\hbox{\sm*}}\fi
\advance\n*1\ifnum\n*<\N*\repeat}}

\def\*shl*#1{\c*=\the\n*\d**\advance\c*#1\a**\d*\dt*\advance\d*#1\b**
\a*=\the\toks0\c*\b*=\the\toks1\d*\advance\a*-\b*
\b*=\the\toks1\c*\d*=\the\toks0\d*\advance\b*\d*
\rx* \the\cos*\a* \tmp* \the\sin*\b* \advance\rx*-\tmp*
\ry* \the\cos*\b* \tmp* \the\sin*\a* \advance\ry*\tmp*
\raise\ry*\rlap{\kern\rx*\unhcopy\spl*}}

\def\calcnormal*#1{\b**=10000sp\a**\b**\k*\n*\advance\k*-\m*
\multiply\a**\k*\divide\a**\m*\a**=#1\a**\ifdim\a**<0pt\a**-\a**\fi
\ifdim\a**>\b**\d*=.96\a**\advance\d*.4\b**
\else\d*=.96\b**\advance\d*.4\a**\fi
\d*=.01\d*\r*\d*\divide\a**\r*\divide\b**\r*
\ifnum\k*<0\a**-\a**\fi\d*=#1\d*\ifdim\d*<0pt\b**-\b**\fi
\k*\a**\a**=\the\k*\dd*\k*\b**\b**=\the\k*\dd*}

\def\wavearcto#1(#2,#3)[#4]{\rlap{\toks0={#2}\toks1={#3}\relax
\calcnum*#1(#2,#3)\c*=\the\N*5sp\a*=.4\c*\a*=#4\a*\ifdim\a*<0pt\a*-\a*\fi
\advance\c*\a*\N*\c*\divide\N*20\multiply\N*2\advance\N*-1\multiply\N*4\relax
\d**=#1\Lengthunit\dd*=.012\d**
\divide\dd*\*ths \multiply\dd*\magnitude
\ifdim\d**<0pt\d**-\d**\fi\L*=.25\d**
\divide\d**\N*\divide\dd*\N*\L*=#4\L*\divide\L*\*ths
\m*\N*\divide\m*2\dm*=\the\m*0sp\l*\dm*
\sm*\n*\*one\loop\calcnormal*{#4}\calcparab*
\*shl*{1}\advance\n*\*one\calcparab*
\*shl*{1.3}\advance\n*\*one\calcparab*
\*shl*{1}\advance\n*2\dd*-\dd*\ifnum\n*<\N*\repeat\n*\N*\shl**{0pt}}}

\def\triangarcto#1(#2,#3)[#4]{\rlap{\toks0={#2}\toks1={#3}\relax
\calcnum*#1(#2,#3)\c*=\the\N*5sp\a*=.4\c*\a*=#4\a*\ifdim\a*<0pt\a*-\a*\fi
\advance\c*\a*\N*\c*\divide\N*20\multiply\N*2\advance\N*-1\multiply\N*2\relax
\d**=#1\Lengthunit\dd*=.012\d**
\divide\dd*\*ths \multiply\dd*\magnitude
\ifdim\d**<0pt\d**-\d**\fi\L*=.25\d**
\divide\d**\N*\divide\dd*\N*\L*=#4\L*\divide\L*\*ths
\m*\N*\divide\m*2\dm*=\the\m*0sp\l*\dm*
\sm*\n*\*one\loop\calcnormal*{#4}\calcparab*
\*shl*{1}\advance\n*2\dd*-\dd*\ifnum\n*<\N*\repeat\n*\N*\shl**{0pt}}}

\def\hr*#1{\L*=\xscale\Lengthunit\ifnum
\angle**=0\clap{\vrule width#1\L* height.1pt}\else
\L*=#1\L*\L*=.5\L*\rmov*(-\L*,0pt){\sm*}\rmov*(\L*,0pt){\sl*}\fi}

\def\shade#1[#2]{\rlap{\Lengthunit=#1\Lengthunit
\special{em:linewidth .001pt}\relax
\mov(0,#2.05){\hr*{.994}}\mov(0,#2.1){\hr*{.980}}\relax
\mov(0,#2.15){\hr*{.953}}\mov(0,#2.2){\hr*{.916}}\relax
\mov(0,#2.25){\hr*{.867}}\mov(0,#2.3){\hr*{.798}}\relax
\mov(0,#2.35){\hr*{.715}}\mov(0,#2.4){\hr*{.603}}\relax
\mov(0,#2.45){\hr*{.435}}\special{em:linewidth \the\linwid*}}}

\def\dshade#1[#2]{\rlap{\special{em:linewidth .001pt}\relax
\Lengthunit=#1\Lengthunit\if#2-\def\t*{+}\else\def\t*{-}\fi
\mov(0,\t*.025){\relax
\mov(0,#2.05){\hr*{.995}}\mov(0,#2.1){\hr*{.988}}\relax
\mov(0,#2.15){\hr*{.969}}\mov(0,#2.2){\hr*{.937}}\relax
\mov(0,#2.25){\hr*{.893}}\mov(0,#2.3){\hr*{.836}}\relax
\mov(0,#2.35){\hr*{.760}}\mov(0,#2.4){\hr*{.662}}\relax
\mov(0,#2.45){\hr*{.531}}\mov(0,#2.5){\hr*{.320}}\relax
\special{em:linewidth \the\linwid*}}}}

\def\vdot{\rlap{\kern-1.9pt\lower1.8pt\hbox{$\scriptstyle\bullet$}}}
\def\vtimes{\rlap{\kern-3pt\lower1.8pt\hbox{$\scriptstyle\times$}}}
\def\vDot{\rlap{\kern-2.3pt\lower2.7pt\hbox{$\bullet$}}}
\def\vTimes{\rlap{\kern-3.6pt\lower2.4pt\hbox{$\times$}}}

\def\arc(#1)[#2,#3]{{\k*=#2\l*=#3\m*=\l*
\advance\m*-6\ifnum\k*>\l*\relax\else
{\rotate(#2)\mov(#1,0){\sm*}}\loop
\ifnum\k*<\m*\advance\k*5{\rotate(\k*)\mov(#1,0){\sl*}}\repeat
{\rotate(#3)\mov(#1,0){\sl*}}\fi}}

\def\dasharc(#1)[#2,#3]{{\k**=#2\n*=#3\advance\n*-1\advance\n*-\k**
\L*=1000sp\L*#1\L* \multiply\L*\n* \multiply\L*\Nhalfperiods
\divide\L*57\N*\L* \divide\N*2000\ifnum\N*=0\N*1\fi
\r*\n*  \divide\r*\N* \ifnum\r*<2\r*2\fi
\m**\r* \divide\m**2 \l**\r* \advance\l**-\m** \N*\n* \divide\N*\r*
\k**\r* \multiply\k**\N* \dn*\n*
\advance\dn*-\k** \divide\dn*2\advance\dn*\*one
\r*\l** \divide\r*2\advance\dn*\r* \advance\N*-2\k**#2\relax
\ifnum\l**<6{\rotate(#2)\mov(#1,0){\sm*}}\advance\k**\dn*
{\rotate(\k**)\mov(#1,0){\sl*}}\advance\k**\m**
{\rotate(\k**)\mov(#1,0){\sm*}}\loop
\advance\k**\l**{\rotate(\k**)\mov(#1,0){\sl*}}\advance\k**\m**
{\rotate(\k**)\mov(#1,0){\sm*}}\advance\N*-1\ifnum\N*>0\repeat
{\rotate(#3)\mov(#1,0){\sl*}}\else\advance\k**\dn*
\arc(#1)[#2,\k**]\loop\advance\k**\m** \r*\k**
\advance\k**\l** {\arc(#1)[\r*,\k**]}\relax
\advance\N*-1\ifnum\N*>0\repeat
\advance\k**\m**\arc(#1)[\k**,#3]\fi}}

\def\triangarc#1(#2)[#3,#4]{{\k**=#3\n*=#4\advance\n*-\k**
\L*=1000sp\L*#2\L* \multiply\L*\n* \multiply\L*\Nhalfperiods
\divide\L*57\N*\L* \divide\N*1000\ifnum\N*=0\N*1\fi
\d**=#2\Lengthunit \d*\d** \divide\d*57\multiply\d*\n*
\r*\n*  \divide\r*\N* \ifnum\r*<2\r*2\fi
\m**\r* \divide\m**2 \l**\r* \advance\l**-\m** \N*\n* \divide\N*\r*
\dt*\d* \divide\dt*\N* \dt*.5\dt* \dt*#1\dt*
\divide\dt*1000\multiply\dt*\magnitude
\k**\r* \multiply\k**\N* \dn*\n* \advance\dn*-\k** \divide\dn*2\relax
\r*\l** \divide\r*2\advance\dn*\r* \advance\N*-1\k**#3\relax
{\rotate(#3)\mov(#2,0){\sm*}}\advance\k**\dn*
{\rotate(\k**)\mov(#2,0){\sl*}}\advance\k**-\m**\advance\l**\m**\loop\dt*-\dt*
\d*\d** \advance\d*\dt*
\advance\k**\l**{\rotate(\k**)\rmov*(\d*,0pt){\sl*}}%
\advance\N*-1\ifnum\N*>0\repeat\advance\k**\m**
{\rotate(\k**)\mov(#2,0){\sl*}}{\rotate(#4)\mov(#2,0){\sl*}}}}

\def\wavearc#1(#2)[#3,#4]{{\k**=#3\n*=#4\advance\n*-\k**
\L*=4000sp\L*#2\L* \multiply\L*\n* \multiply\L*\Nhalfperiods
\divide\L*57\N*\L* \divide\N*1000\ifnum\N*=0\N*1\fi
\d**=#2\Lengthunit \d*\d** \divide\d*57\multiply\d*\n*
\r*\n*  \divide\r*\N* \ifnum\r*=0\r*1\fi
\m**\r* \divide\m**2 \l**\r* \advance\l**-\m** \N*\n* \divide\N*\r*
\dt*\d* \divide\dt*\N* \dt*.7\dt* \dt*#1\dt*
\divide\dt*1000\multiply\dt*\magnitude
\k**\r* \multiply\k**\N* \dn*\n* \advance\dn*-\k** \divide\dn*2\relax
\divide\N*4\advance\N*-1\k**#3\relax
{\rotate(#3)\mov(#2,0){\sm*}}\advance\k**\dn*
{\rotate(\k**)\mov(#2,0){\sl*}}\advance\k**-\m**\advance\l**\m**\loop\dt*-\dt*
\d*\d** \advance\d*\dt* \dd*\d** \advance\dd*1.3\dt*
\advance\k**\r*{\rotate(\k**)\rmov*(\d*,0pt){\sl*}}\relax
\advance\k**\r*{\rotate(\k**)\rmov*(\dd*,0pt){\sl*}}\relax
\advance\k**\r*{\rotate(\k**)\rmov*(\d*,0pt){\sl*}}\relax
\advance\k**\r*
\advance\N*-1\ifnum\N*>0\repeat\advance\k**\m**
{\rotate(\k**)\mov(#2,0){\sl*}}{\rotate(#4)\mov(#2,0){\sl*}}}}

\def\gmov*#1(#2,#3)#4{\rlap{\L*=#1\Lengthunit
\xL*=#2\L* \yL*=#3\L*
\rx* \gcos*\xL* \tmp* \gsin*\yL* \advance\rx*-\tmp*
\ry* \gcos*\yL* \tmp* \gsin*\xL* \advance\ry*\tmp*
\rx*=\xscale\rx* \ry*=\yscale\ry*
\xL* \the\cos*\rx* \tmp* \the\sin*\ry* \advance\xL*-\tmp*
\yL* \the\cos*\ry* \tmp* \the\sin*\rx* \advance\yL*\tmp*
\kern\xL*\raise\yL*\hbox{#4}}}

\def\rgmov*(#1,#2)#3{\rlap{\xL*#1\yL*#2\relax
\rx* \gcos*\xL* \tmp* \gsin*\yL* \advance\rx*-\tmp*
\ry* \gcos*\yL* \tmp* \gsin*\xL* \advance\ry*\tmp*
\rx*=\xscale\rx* \ry*=\yscale\ry*
\xL* \the\cos*\rx* \tmp* \the\sin*\ry* \advance\xL*-\tmp*
\yL* \the\cos*\ry* \tmp* \the\sin*\rx* \advance\yL*\tmp*
\kern\xL*\raise\yL*\hbox{#3}}}

\def\Earc(#1)[#2,#3][#4,#5]{{\k*=#2\l*=#3\m*=\l*
\advance\m*-6\ifnum\k*>\l*\relax\else\def\xscale{#4}\def\yscale{#5}\relax
{\angle**0\rotate(#2)}\gmov*(#1,0){\sm*}\loop
\ifnum\k*<\m*\advance\k*5\relax
{\angle**0\rotate(\k*)}\gmov*(#1,0){\sl*}\repeat
{\angle**0\rotate(#3)}\gmov*(#1,0){\sl*}\relax
\def\xscale{1}\def\yscale{1}\fi}}

\def\dashEarc(#1)[#2,#3][#4,#5]{{\k**=#2\n*=#3\advance\n*-1\advance\n*-\k**
\L*=1000sp\L*#1\L* \multiply\L*\n* \multiply\L*\Nhalfperiods
\divide\L*57\N*\L* \divide\N*2000\ifnum\N*=0\N*1\fi
\r*\n*  \divide\r*\N* \ifnum\r*<2\r*2\fi
\m**\r* \divide\m**2 \l**\r* \advance\l**-\m** \N*\n* \divide\N*\r*
\k**\r*\multiply\k**\N* \dn*\n* \advance\dn*-\k** \divide\dn*2\advance\dn*\*one
\r*\l** \divide\r*2\advance\dn*\r* \advance\N*-2\k**#2\relax
\ifnum\l**<6\def\xscale{#4}\def\yscale{#5}\relax
{\angle**0\rotate(#2)}\gmov*(#1,0){\sm*}\advance\k**\dn*
{\angle**0\rotate(\k**)}\gmov*(#1,0){\sl*}\advance\k**\m**
{\angle**0\rotate(\k**)}\gmov*(#1,0){\sm*}\loop
\advance\k**\l**{\angle**0\rotate(\k**)}\gmov*(#1,0){\sl*}\advance\k**\m**
{\angle**0\rotate(\k**)}\gmov*(#1,0){\sm*}\advance\N*-1\ifnum\N*>0\repeat
{\angle**0\rotate(#3)}\gmov*(#1,0){\sl*}\def\xscale{1}\def\yscale{1}\else
\advance\k**\dn* \Earc(#1)[#2,\k**][#4,#5]\loop\advance\k**\m** \r*\k**
\advance\k**\l** {\Earc(#1)[\r*,\k**][#4,#5]}\relax
\advance\N*-1\ifnum\N*>0\repeat
\advance\k**\m**\Earc(#1)[\k**,#3][#4,#5]\fi}}

\def\triangEarc#1(#2)[#3,#4][#5,#6]{{\k**=#3\n*=#4\advance\n*-\k**
\L*=1000sp\L*#2\L* \multiply\L*\n* \multiply\L*\Nhalfperiods
\divide\L*57\N*\L* \divide\N*1000\ifnum\N*=0\N*1\fi
\d**=#2\Lengthunit \d*\d** \divide\d*57\multiply\d*\n*
\r*\n*  \divide\r*\N* \ifnum\r*<2\r*2\fi
\m**\r* \divide\m**2 \l**\r* \advance\l**-\m** \N*\n* \divide\N*\r*
\dt*\d* \divide\dt*\N* \dt*.5\dt* \dt*#1\dt*
\divide\dt*1000\multiply\dt*\magnitude
\k**\r* \multiply\k**\N* \dn*\n* \advance\dn*-\k** \divide\dn*2\relax
\r*\l** \divide\r*2\advance\dn*\r* \advance\N*-1\k**#3\relax
\def\xscale{#5}\def\yscale{#6}\relax
{\angle**0\rotate(#3)}\gmov*(#2,0){\sm*}\advance\k**\dn*
{\angle**0\rotate(\k**)}\gmov*(#2,0){\sl*}\advance\k**-\m**
\advance\l**\m**\loop\dt*-\dt* \d*\d** \advance\d*\dt*
\advance\k**\l**{\angle**0\rotate(\k**)}\rgmov*(\d*,0pt){\sl*}\relax
\advance\N*-1\ifnum\N*>0\repeat\advance\k**\m**
{\angle**0\rotate(\k**)}\gmov*(#2,0){\sl*}\relax
{\angle**0\rotate(#4)}\gmov*(#2,0){\sl*}\def\xscale{1}\def\yscale{1}}}

\def\waveEarc#1(#2)[#3,#4][#5,#6]{{\k**=#3\n*=#4\advance\n*-\k**
\L*=4000sp\L*#2\L* \multiply\L*\n* \multiply\L*\Nhalfperiods
\divide\L*57\N*\L* \divide\N*1000\ifnum\N*=0\N*1\fi
\d**=#2\Lengthunit \d*\d** \divide\d*57\multiply\d*\n*
\r*\n*  \divide\r*\N* \ifnum\r*=0\r*1\fi
\m**\r* \divide\m**2 \l**\r* \advance\l**-\m** \N*\n* \divide\N*\r*
\dt*\d* \divide\dt*\N* \dt*.7\dt* \dt*#1\dt*
\divide\dt*1000\multiply\dt*\magnitude
\k**\r* \multiply\k**\N* \dn*\n* \advance\dn*-\k** \divide\dn*2\relax
\divide\N*4\advance\N*-1\k**#3\def\xscale{#5}\def\yscale{#6}\relax
{\angle**0\rotate(#3)}\gmov*(#2,0){\sm*}\advance\k**\dn*
{\angle**0\rotate(\k**)}\gmov*(#2,0){\sl*}\advance\k**-\m**
\advance\l**\m**\loop\dt*-\dt*
\d*\d** \advance\d*\dt* \dd*\d** \advance\dd*1.3\dt*
\advance\k**\r*{\angle**0\rotate(\k**)}\rgmov*(\d*,0pt){\sl*}\relax
\advance\k**\r*{\angle**0\rotate(\k**)}\rgmov*(\dd*,0pt){\sl*}\relax
\advance\k**\r*{\angle**0\rotate(\k**)}\rgmov*(\d*,0pt){\sl*}\relax
\advance\k**\r*
\advance\N*-1\ifnum\N*>0\repeat\advance\k**\m**
{\angle**0\rotate(\k**)}\gmov*(#2,0){\sl*}\relax
{\angle**0\rotate(#4)}\gmov*(#2,0){\sl*}\def\xscale{1}\def\yscale{1}}}

\newcount\CatcodeOfAtSign
\CatcodeOfAtSign=\the\catcode`\@
\catcode`\@=11
\def\@arc#1[#2][#3]{\rlap{\Lengthunit=#1\Lengthunit
\sm*\l*arc(#2.1914,#3.0381)[#2][#3]\relax
\mov(#2.1914,#3.0381){\l*arc(#2.1622,#3.1084)[#2][#3]}\relax
\mov(#2.3536,#3.1465){\l*arc(#2.1084,#3.1622)[#2][#3]}\relax
\mov(#2.4619,#3.3086){\l*arc(#2.0381,#3.1914)[#2][#3]}}}

\def\dash@arc#1[#2][#3]{\rlap{\Lengthunit=#1\Lengthunit
\d*arc(#2.1914,#3.0381)[#2][#3]\relax
\mov(#2.1914,#3.0381){\d*arc(#2.1622,#3.1084)[#2][#3]}\relax
\mov(#2.3536,#3.1465){\d*arc(#2.1084,#3.1622)[#2][#3]}\relax
\mov(#2.4619,#3.3086){\d*arc(#2.0381,#3.1914)[#2][#3]}}}

\def\wave@arc#1[#2][#3]{\rlap{\Lengthunit=#1\Lengthunit
\w*lin(#2.1914,#3.0381)\relax
\mov(#2.1914,#3.0381){\w*lin(#2.1622,#3.1084)}\relax
\mov(#2.3536,#3.1465){\w*lin(#2.1084,#3.1622)}\relax
\mov(#2.4619,#3.3086){\w*lin(#2.0381,#3.1914)}}}

\def\bezier#1(#2,#3)(#4,#5)(#6,#7){\N*#1\l*\N* \advance\l*\*one
\d* #4\Lengthunit \advance\d* -#2\Lengthunit \multiply\d* \*two
\b* #6\Lengthunit \advance\b* -#2\Lengthunit
\advance\b*-\d* \divide\b*\N*
\d** #5\Lengthunit \advance\d** -#3\Lengthunit \multiply\d** \*two
\b** #7\Lengthunit \advance\b** -#3\Lengthunit
\advance\b** -\d** \divide\b**\N*
\mov(#2,#3){\sm*{\loop\ifnum\m*<\l*
\a*\m*\b* \advance\a*\d* \divide\a*\N* \multiply\a*\m*
\a**\m*\b** \advance\a**\d** \divide\a**\N* \multiply\a**\m*
\rmov*(\a*,\a**){\unhcopy\spl*}\advance\m*\*one\repeat}}}

\catcode`\*=12

\newcount\n@ast

\def\n@ast@#1{\n@ast0\relax\get@ast@#1\end}
\def\get@ast@#1{\ifx#1\end\let\next\relax\else
\ifx#1*\advance\n@ast1\fi\let\next\get@ast@\fi\next}

\newif\if@up \newif\if@dwn
\def\up@down@#1{\@upfalse\@dwnfalse
\if#1u\@uptrue\fi\if#1U\@uptrue\fi\if#1+\@uptrue\fi
\if#1d\@dwntrue\fi\if#1D\@dwntrue\fi\if#1-\@dwntrue\fi}

\def\halfcirc#1(#2)[#3]{{\Lengthunit=#2\Lengthunit\up@down@{#3}\relax
\if@up\mov(0,.5){\@arc[-][-]\@arc[+][-]}\fi
\if@dwn\mov(0,-.5){\@arc[-][+]\@arc[+][+]}\fi
\def\lft{\mov(0,.5){\@arc[-][-]}\mov(0,-.5){\@arc[-][+]}}\relax
\def\rght{\mov(0,.5){\@arc[+][-]}\mov(0,-.5){\@arc[+][+]}}\relax
\if#3l\lft\fi\if#3L\lft\fi\if#3r\rght\fi\if#3R\rght\fi
\n@ast@{#1}\relax
\ifnum\n@ast>0\if@up\shade[+]\fi\if@dwn\shade[-]\fi\fi
\ifnum\n@ast>1\if@up\dshade[+]\fi\if@dwn\dshade[-]\fi\fi}}

\def\halfdashcirc(#1)[#2]{{\Lengthunit=#1\Lengthunit\up@down@{#2}\relax
\if@up\mov(0,.5){\dash@arc[-][-]\dash@arc[+][-]}\fi
\if@dwn\mov(0,-.5){\dash@arc[-][+]\dash@arc[+][+]}\fi
\def\lft{\mov(0,.5){\dash@arc[-][-]}\mov(0,-.5){\dash@arc[-][+]}}\relax
\def\rght{\mov(0,.5){\dash@arc[+][-]}\mov(0,-.5){\dash@arc[+][+]}}\relax
\if#2l\lft\fi\if#2L\lft\fi\if#2r\rght\fi\if#2R\rght\fi}}

\def\halfwavecirc(#1)[#2]{{\Lengthunit=#1\Lengthunit\up@down@{#2}\relax
\if@up\mov(0,.5){\wave@arc[-][-]\wave@arc[+][-]}\fi
\if@dwn\mov(0,-.5){\wave@arc[-][+]\wave@arc[+][+]}\fi
\def\lft{\mov(0,.5){\wave@arc[-][-]}\mov(0,-.5){\wave@arc[-][+]}}\relax
\def\rght{\mov(0,.5){\wave@arc[+][-]}\mov(0,-.5){\wave@arc[+][+]}}\relax
\if#2l\lft\fi\if#2L\lft\fi\if#2r\rght\fi\if#2R\rght\fi}}

\catcode`\*=11

\def\Circle#1(#2){\halfcirc#1(#2)[u]\halfcirc#1(#2)[d]\n@ast@{#1}\relax
\ifnum\n@ast>0\L*=\xscale\Lengthunit
\ifnum\angle**=0\clap{\vrule width#2\L* height.1pt}\else
\L*=#2\L*\L*=.5\L*\special{em:linewidth .001pt}\relax
\rmov*(-\L*,0pt){\sm*}\rmov*(\L*,0pt){\sl*}\relax
\special{em:linewidth \the\linwid*}\fi\fi}

\catcode`\*=12

\def\wavecirc(#1){\halfwavecirc(#1)[u]\halfwavecirc(#1)[d]}
\def\dashcirc(#1){\halfdashcirc(#1)[u]\halfdashcirc(#1)[d]}

\def\xscale{1}

\def\yscale{1}

\def\Ellipse#1(#2)[#3,#4]{\def\xscale{#3}\def\yscale{#4}\relax
\Circle#1(#2)\def\xscale{1}\def\yscale{1}}

\def\dashEllipse(#1)[#2,#3]{\def\xscale{#2}\def\yscale{#3}\relax
\dashcirc(#1)\def\xscale{1}\def\yscale{1}}

\def\waveEllipse(#1)[#2,#3]{\def\xscale{#2}\def\yscale{#3}\relax
\wavecirc(#1)\def\xscale{1}\def\yscale{1}}

\def\halfEllipse#1(#2)[#3][#4,#5]{\def\xscale{#4}\def\yscale{#5}\relax
\halfcirc#1(#2)[#3]\def\xscale{1}\def\yscale{1}}

\def\halfdashEllipse(#1)[#2][#3,#4]{\def\xscale{#3}\def\yscale{#4}\relax
\halfdashcirc(#1)[#2]\def\xscale{1}\def\yscale{1}}

\def\halfwaveEllipse(#1)[#2][#3,#4]{\def\xscale{#3}\def\yscale{#4}\relax
\halfwavecirc(#1)[#2]\def\xscale{1}\def\yscale{1}}

\catcode`\@=\the\CatcodeOfAtSign

\title{\boldmath Large $N$ limit of supersymmetric Chern-Simons-matter model: Breakdown of superconformal symmetry}

\author{J. M. Queiruga}
\affiliation{Instituto de F\'\i sica, Universidade de S\~ao Paulo\\
Caixa Postal 66318, 05315-970, S\~ao Paulo, SP, Brazil}
\email{queiruga@if.usp.br}

\author{ A. J. da Silva}

\affiliation{Instituto de F\'\i sica, Universidade de S\~ao Paulo\\
Caixa Postal 66318, 05315-970, S\~ao Paulo, SP, Brazil}
\email{ajsilva@if.usp.br}

\begin{abstract}
We study some properties of the non-abelian, classically conformally invariant, three-dimensional $U(N)$ supersymetric Chern-Simons, coupled to a scalar superfield in the fundamental representation of $U(N)$, in the large $N$ limit. In leading order in $1/N$ we show that the theory has two phases: one in which it remains conformally invariant, and other where the superconformal symmetry is broken and masses for the matter fields are generated.
\end{abstract}

\maketitle

\section{Introduction}

The AdS/CFT correspondence \cite{Maldacena} is an exact duality between quantum theory of gravity containing the anti-de Sitter space AdS$_{d+1}$ and conformal field theories in $d$ dimensions. Despite the fact that we know how to translate the calculation from one side to the other in the correspondence, the most difficult point is to find which quantum gravity theory is dual to the corresponding conformal field theory, since one or both of them could be strongly coupled. The large $N$ limit of $O(N)$ and $U(N)$ Chern-Simons theory coupled to scalar fields in the fundamental representation is conjectured to be dual to Vasiliev's higher spin gravity theory on AdS$_4$ \cite{Klebanov,Fradkin},  and in this case both sides of the correspondence are weakly coupled. This fact has attracted the attention on the large $N$ limit of Chern-Simons theories \cite{Giombi,Aharony,Sachin,Aharony1,Yacoby,Maldacena1,Jain,Minwalla}, coupled to matter fields, both scalars or fermions. More recently, spontaneous breaking of the conformal symmetry was studied in different models containing a nonsupersymmetric Chern-Simons term  \cite{Bardeen, Moshe, Bardeen1}, and in a supersymmetric (SUSY) version for a truncated large $N$ limit or perturbative expansions in \cite{Lehum, Lehum1,Ferrari}.

In this work we study the possibility of a dynamical breaking of the superconformal symmetry in a SUSY ($\mathcal{N}=1$) non-abelian Chern-Simons theory coupled to scalar superfields. We work directly in superfield formalism, which means that each supergraph contains all possible contributions of the component fields when we integrate the Grassmann coordinates.

This work is organized as follows:
In Sec. II the model is presented in terms of superfields and following the methods of \cite{Coleman} in its supersymmetric version, several components of the fields are shifted in a classical background superfield and a quantum part. After obtaining the leading contribution for the classical action in the $1/N$ expansion, in Sec. III we determine the one-loop part, obtaining a surprisingly simple result in such limit, similar to the one obtained in \cite{Kang} for a nonsupersymmetric electrodynamics.
In Sec. IV, following standard methods of $D$ algebra, \cite{Wess, Buchbinder}, we write down the superfield propagators and we determine the remaining contributions at leading order in $N$ and up to order $g^2$. We obtain  the effective potential, which is exact in $\lambda$ (the marginal coupling constant) and up to order $\mathcal{O}(g^2)$  and leading order in $N$. 
In Sec. V we solve the ``gap equations", analyzing the possibility of dynamical breaking of superconformal symmetry, finding a nonbreaking phase, where no masses are generated for any of the fields, and a massive phase, where superconformal symmetry is broken. Finally the last section is devoted to the discussion and conclusions.

%%%%%%%%%%%%%%%%%
%%%%%%%%%%%%%%%%%
%%%%%%%%%%%%%%%%%
%%%%%%%%%%%%%%%%%

\section{The $\mathcal{N}=1$ SUSY Chern-Simons-matter model}

The $\mathcal{N}=1$ three-dimensional $U(N)$ SUSY Chern-Simons (SCS) model is defined by the classical action (see \cite{GrisaruB}):
\begin{eqnarray}
S_{CS}&=&\int d^5z\, tr \left\{\Gamma^\alpha W_\alpha+\frac{ig}{6\sqrt{N}}\lbrace \Gamma^\alpha,\Gamma^\beta\rbrace D_\beta \Gamma_\alpha+\frac{g^2}{12 N}\lbrace \Gamma^\alpha,\Gamma^\beta\rbrace\lbrace \Gamma_\alpha,\Gamma_\beta\rbrace    \right\}\nonumber\\
&=&\int d^5z\, tr \left\{-\frac{1}{2}\Gamma_{\alpha}D^{\beta}D^{\alpha}\Gamma_{\beta}-\frac{ig}{3 \sqrt{N}}\Gamma^{\alpha}\Gamma^{\beta}D_{\beta}\Gamma_{\alpha}\right. \nonumber\\
&&\left. -\frac{ig}{3\sqrt{N}}\Gamma^{\alpha}\Gamma^{\beta}D_{\alpha}\Gamma_{\beta}
-\frac{g^2}{6N}\Gamma^{\alpha}\Gamma_{\alpha}\Gamma^{\beta}\Gamma_{\beta}
-\frac{g^2}{6N}\Gamma^{\alpha}\Gamma^{\beta}\Gamma_{\alpha}\Gamma_{\beta}\right\}
\label{CS}
\end{eqnarray}
where the fields and notations are given in Eqs. (\ref{x1})-(\ref{x2}), below. We are interested in the study of the possible superconformal invariance breaking and mass generation of the SCS interacting with a massless and self -interacting matter field, so we introduce the following matter Lagrangian:
\be
S_{mat}=\int d^5 z \left\{-\frac{1}{2}(\nabla^\alpha \Phi)^\dagger (\nabla_\alpha\Phi)+\frac{\lambda}{2N}(\Phi^\dagger\Phi)^2\right\}
\label{mat}
\ee
where:
\begin{eqnarray}
\nabla^\alpha&=&D^\alpha-i\frac{g}{\sqrt{N}}\Gamma^\alpha,\,\,\,\,\,\,\,\  D_{\alpha}= \partial_{\alpha} +i \theta^{\beta} \partial_{\beta \alpha}\,\,\,\,\,\,\,\,\alpha, \beta=1,2\label{x1}\\
W_\alpha&=&\frac{1}{2}D^\beta D_\alpha \Gamma_\beta-\frac{ig}{2\sqrt{N}}\lbrack  \Gamma^\beta,D_\beta \Gamma_\alpha \rbrack-\frac{g}{6N}\lbrack \Gamma^\beta,\lbrace \Gamma_\beta,\Gamma_\alpha\rbrace\rbrack\\
\Gamma^{\alpha}&=&\chi^{\alpha}-\theta^{\alpha}B-i\theta_{\beta} A^{\beta \alpha}-\theta^2 (2 \rho^{\alpha}-i\partial^{\alpha \beta} \chi_{\beta})\\
\Gamma^\alpha&=&\Gamma^\alpha_A T_A,\quad T_A \in u(N) \,\,\,\,\,\,\,\,\,\,\,\,\, A=1,2\cdots N^2 \\
\Phi^a&=& \phi^a+ \theta^{\alpha}\psi_{\alpha}^a -F^a \theta^2 \quad\quad\quad a=1,2\cdots N\label{x2}
\end{eqnarray}

Our metric is $g_{\mu \nu}=diag (-, +,+)$ and the spinorial indices $(\alpha, \beta=1,2)$ are raised and lowered by  $C_{\alpha\beta}=-C^{\alpha\beta}=\tau_2$ (the second Pauli matrix), to know: $\Gamma^{\alpha}=C^{\alpha \beta} \Gamma_{\beta}$ and  $\Gamma_{\alpha}=\Gamma^{\beta} C_{\beta \alpha}$. 
The spinorial derivative $\partial_{\alpha}$ is defined by  
$\partial_{\alpha}=\frac{\partial}{\partial \theta^{\alpha}}$, and $\theta^2=\frac{1}{2} \theta^{\alpha} \theta_{\alpha}$.

The spinorial gauge superfield $\Gamma^{\alpha}$ is in the adjoint representation of the group and the scalar matter superfield $\Phi=[\phi^a]$ with $a=1,2...N$, is in the fundamental representation. The spinorial superfield $\Gamma_{\alpha}$ is composed, in the Wess-Zumino gauge, by the gauge potential $A^\mu=-\frac{1}{2}(\gamma^{\mu})_{\alpha \beta} 
A^{\alpha \beta}$ (where $\gamma^\mu$ are Dirac matrices, $\alpha,\beta=1,2$ are spinorial indices and $\mu, \nu=0,1,2$ are space-time indices) and the gaugino $\rho_{\alpha}$. In a SUSY covariant gauge (in which we will work) it has yet the auxiliary fields $\chi^{\alpha}$ and $B$. The vector superfield $\Phi_a$ is composed by the scalar matter field $\phi_{a}$, the spinorial field $\psi_a^{\alpha}$ and the auxiliary field $F_{a}$. 

\begin{figure}[h]
    \centering
    \includegraphics[width=0.5\textwidth]{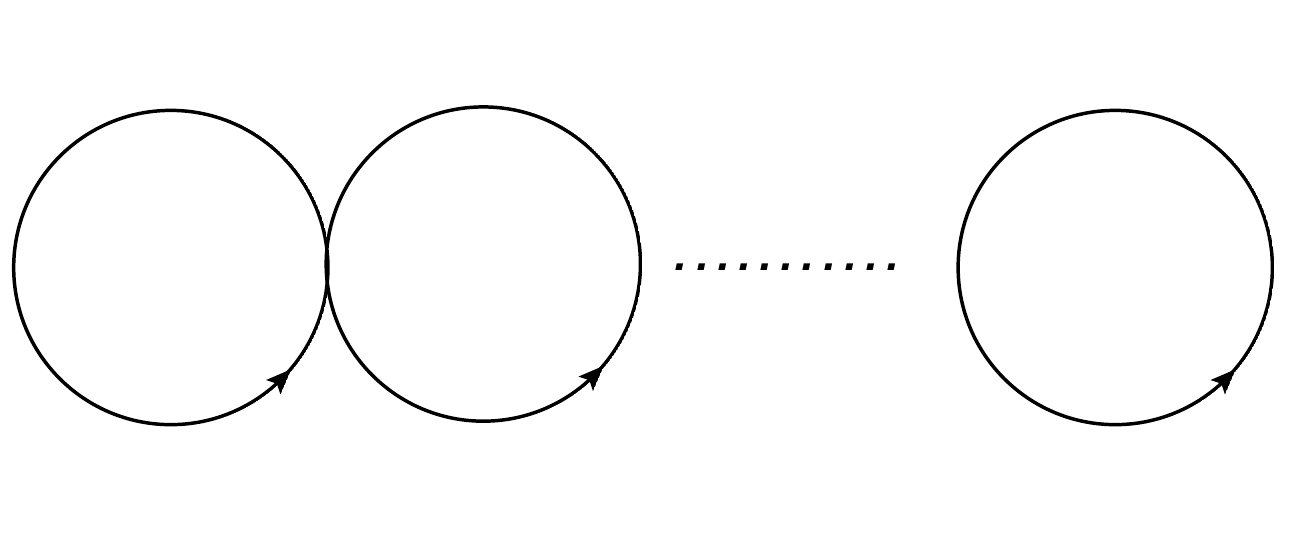}
    \caption{Vacuum bubbles contributing at order N}
    \label{bubble}
\end{figure}

The two parameters $g$ and $\lambda$ are dimensionless and the model is classically conformally invariant. To favor the study of the model in the $1/N$ expansion, we introduce, in the way of Coleman \textit{et al.} \cite{Coleman}, the extra term

\be
S_{aux}=-\int d^5z \frac{1}{2}\left\{ \Sigma -\sqrt{\frac{\lambda}{N}}\Phi^\dagger\Phi        \right\}^2 \label{saux}
\ee
where $\Sigma$ is a real, scalar, $U(N)$ singlet superfield. This added term does not affect the dynamics of the original theory, since after functionally integrating over $\Sigma$ (a trivial Gaussian integral), it gives an irrelevant constant multiplying the original generating functional. Note that (\ref{saux}) eliminates the quartic term in (\ref{mat}). The consequence of this is the reduction of the infinite number of diagrams contributing at leading order and involving $\Phi^a$ loops (Fig. \ref{bubble}) to a single one-loop diagram.

To fix the gauge, we introduce the following (SUSY covariant) gauge fixing and Faddeev-Popov (FP) ghost actions
\be
S_{gf}=-\int d^5z \frac{1}{2\alpha}tr\left\{D^\alpha \Gamma_\alpha D^\beta \Gamma_\beta   \right\}
\ee
\be
S_{FP}=\int d^5 z\, tr \left\{ c^\dagger D^2 c -i\frac{g}{2\sqrt{N}}c^\dagger \lbrack D^\alpha\Gamma_\alpha,c\rbrack-i\frac{g}{2\sqrt{N}}c^\dagger \lbrack \Gamma^\alpha,D_\alpha c \rbrack \right\},   
\ee
where the FP ghost fields are in adjoint representation of the group $c=c^A T^A$ with $A=1,\cdots, N^2$.

The effective potential defined by $V_{eff}(b_{c},\sigma_{c})\equiv -(1/L^3) S_{eff}(b_{c},\sigma_{c})$, where $S_{eff}$ is the effective action for classical constant fields, and $L^3$ is the volume of the space-time, can be calculated by the functional method of Jackiw \cite{Jackiw} (see also \cite{Root}). This requires us to shift the superfields as follows:
\bea
\phi^1&\rightarrow& \varphi+\sqrt{\frac{N}{2}}b_c\label{shiftphi}\\
\Sigma &\rightarrow& \Sigma+\sqrt{\frac{N}{\lambda}}\sigma_c\\
\phi^k&\rightarrow& \phi^k\,\,\,\,\,\,\,\,\,\,\,\,\,\,\,k=2,3\cdots N\\
\Gamma^{\alpha}&\rightarrow&\Gamma^{\alpha}\label{shiftgamma},
\eea
with $b_{c}$ and $\sigma_{c}$ being real constant (in $x^\mu$) classical background superfields: $b_{c}=b_1-\theta^2b_2$ and $\sigma_{c}=\sigma_1-\theta^2\sigma_2$. $\Gamma$ and $\Sigma$ are Hermitian quantum fields and $\varphi$ and $\phi^k$ are complex quantum fields chosen to have zero expectation value, at any order of approximation. From the effective potential $V_{eff}(b_{c}, \sigma_{c})$ obtained by this method, the  potential $V_{eff}(b_{c})$ of the original theory can be obtained by solving the auxiliary field equation of motion: $\partial V_{eff}/\partial \sigma_{c}=0$. 

The calculation of these effective potentials, by the functional method, requires the shift of the quantum fields by their possible non-null, classical expectation values, including components in the direction $\theta^2$ ($\sigma_2$ and $b_2$, in our case). These components explicitly break supersymmetry which makes the calculations with superfields very involved. A formalism of superfields in the presence of broken SUSY, for 2+1 D, was developed in \cite{Helayel} for scalar fields and extended to gauge fields in \cite{Gallegos}. One example of calculation using this method, can be seen in \cite{Maluf}, for the 2+1 D Wess-Zumino model. Happily, in order to study the possibility of conformal symmetry breaking and mass generation, it is enough to calculate the effective potential up to linear dependence in the $\theta^2$ components ($\sigma_2$ and $b_2$, in the present paper)  \cite{AlvarezG, Burguess,Ferrari}. The result obtained in this way is called the K\"ahler effective potential \cite{Buchbinder}. 
In superfield formalism, this approximation can be achieved by throwing away terms in $D_\alpha b_{c}$ and $D_\alpha \sigma_{c}$ \cite{Ferrari1}, in the calculation of the radiative corrections, which means 
to use the rules: $D_{\alpha}\sigma_{c}=\sigma_{c} D_{\alpha}$ and $D_{\alpha}b_{c}=b_{c}D_{\alpha}$, even if not taking $b_2$ and $\sigma_2$ equal to zero in $b_{c}$ and $\sigma_{c}$.

An observation is in order. For non-Abelian gauge theories, the  number of Feynman graphs involved in the leading order of $1/N$, in the $R_\xi$ gauges is infinity (all the planar diagrams), as first advanced by 't Hooft \cite{Hooft}.
So, by following Kang \cite{Kang}, we will consider the extra approximation $g<<1$, stopping the calculations at order $g^2$ (no restriction is needed with respect to the order of the self-coupling constant $\lambda$). In this approximation we will have contribution of diagrams  until two loops. For light-cone gauge calculations see for example \cite{Jain, Minwalla, QS}.

After shifting the fields as in Eqs. (\ref{shiftphi})-(\ref{shiftgamma})  the action results in the sum of
\noindent
(i) the classical term
\be 
\Gamma_{cl}=N \int d^5 z \left(-\frac{1}{4} D^{\alpha} b_{c}D_{\alpha} b_{c}
+\frac{1}{2}\sigma_{c} b_{c}^2 -\frac{1}{2 \lambda}\sigma_{c}^2  \right),
\label{V0}
\ee
(ii) the quadratic part (in the quantum fields) given by
\bea
S_{2}&=&\int d^5 z \left \{(\Phi^a)^\dagger (D^2+\sigma_c)\Phi^a+ \frac{1}{2}\Gamma^\alpha_{11}(\Theta^{\alpha \beta}+\frac{g^2 b_c^2}{2}C^{\alpha \beta} )\Gamma^\beta_{11}+\Gamma^{*\alpha}_{1j}\left( \Theta_{\alpha\beta}+\frac{g^2b_c^2}{4}C_{\alpha\beta}  \right)\Gamma^\beta_{1j} \right.	\nonumber\\
&&\left. +\frac{1}{2}\Gamma^{\alpha}_{ji} \Theta_{\alpha\beta}\Gamma^\beta_{ij}+ i \frac{g}{\sqrt{2}} b_c\Gamma^\alpha_{11}\left(D_\alpha \varphi-D_\alpha \varphi^\dagger\right)+i\frac{g}{2\sqrt{2}}\left(\Gamma^\alpha_{1j}D_\alpha\phi^j -\Gamma^{*\alpha}_{1j}D_\alpha (\phi^j)^\dagger\right) \label{quad}\right.\label{S2}\\
&&\left.+ \sqrt{\frac{\lambda}{2}}\,
b_c \Sigma \left( \varphi+\varphi^\dagger  \right)-\frac{1}{2}\Sigma^2+c^\dagger D^2 c\right\}, \nonumber
\eea
where $a=1,...,N$ and $i,j,k=2,...,N$. We also wrote $\Gamma^\alpha_{j1}=\Gamma^{*\alpha }_{1j}$ in convenient places.\\ 
(iii) the interaction trilinear terms
\bea
S_{3}&=&\int d^5 z \left \{-i\frac{g}{2\sqrt{N}}\left( (\Phi^i)^\dagger\Gamma_{ij}^\alpha D_\alpha \Phi^j-D^\alpha (\Phi^i)^\dagger \Gamma_{\alpha,ji} \Phi^j\right)+\sqrt{\frac{\lambda}{N}}\Sigma (\Phi^i)^\dagger\Phi^i \right.\nonumber\\
&&\left.  + \sqrt{\frac{\lambda}{N}}\Sigma \varphi^\dagger\varphi   -i\frac{g}{3\sqrt{N}}\Gamma^\alpha\Gamma^\beta D_\beta \Gamma_\alpha+i\frac{2g}{3\sqrt{N}}\Gamma^\alpha \Gamma^\beta D_\alpha \Gamma_\beta-i\frac{g}{2\sqrt{N}}c^\dagger \lbrack D^\alpha \Gamma_\alpha,c\rbrack  \right.\\
&&\left.-i\frac{g}{2\sqrt{N}}c^\dagger\lbrack  \Gamma^\alpha,D_\alpha c  \rbrack\right\},\nonumber
\eea
and (iv) the quadrilinear terms
\be
S_{4}=\int d^5 z \left \{ -\frac{g^2}{6
N}\Gamma^\alpha\Gamma_\alpha\Gamma^\beta\Gamma_\beta-\frac{g^2}{6N}\Gamma^\alpha\Gamma^\beta \Gamma_\alpha \Gamma_\beta+\frac{g^2}{2N}\Gamma^\alpha\Gamma_\alpha(\Phi^i)^\dagger\Phi^i+\frac{g^2}{2N}\Gamma^\alpha\Gamma_\alpha\varphi^\dagger\varphi \right\}.
\ee

An action linear in the quantum fields, not involved in the calculations, was omitted.
For later use we define 
\bea
\mathcal{O}&=& D^2+\sigma_c\\
\Theta^{\alpha\beta}&=&-D^\beta D^\alpha-\frac{D^\alpha D^\beta}{\alpha}\\
\Pi^{\alpha\beta}&=&\Theta^{\alpha\beta}+\frac{g^2 b_c^2}{4} C^{\alpha\beta}\\
\tilde \Pi^{\alpha\beta}&=&\Theta^{\alpha\beta}+\frac{g^2 b_c^2}{2} C^{\alpha\beta}
\eea

For future use, we must observe that by integrating the quadratic Lagrangian $S_2$, in the anticommuting dimensions $\theta$ we can verify that the $\phi^1$ fermionic and bosonic component fields  have mass parameters $m_F^2= (\lambda \sigma_1)^2$ and $m_B^2=(\lambda \sigma_1)^2 - \lambda \sigma_2$, respectively.

\section{One-loop contributions to the K\"ahlerian effective potential}

From the expression (\ref{quad}) we can read directly the inverse propagator matrices for the superfields. The first one is given by
\be
\frac{1}{2}\Gamma^\alpha_{ij}\Theta_{\alpha\beta}\delta_{jk}\delta_{il} \Gamma^\beta_{kl}\nonumber
\ee
where $i,j\ge 2$. The corresponding one-loop contribution to the effective action will be (the minus sign comes from the integration in the fermionic fields $\Gamma$)
\be
S^{\Gamma}_1=-i \frac{(N-1)^2}{2}\log \det \Theta_{\alpha\beta},
\ee
which is a term of order $N^2$, but independent of the background fields, and therefore an irrelevant additive constant  contribution to the effective potential. Another $N^2$ order term is given by the one-loop contribution of the ghost fields:
\be
S_{1lp}^C=-i N^2\log\det D^2,
\ee
which is again an irrelevant constant. The following term is of order $N$ and mixes the fields $\Gamma$ and $\Phi$,
\be
\left(\begin{matrix}(\Phi^i)^\dagger  & \Gamma^{*\alpha}_{j1}\end{matrix} \right)\left(\begin{matrix}\delta_{ik}\mathcal{O} & \frac{i}{2\sqrt{2}} g b_c \delta_{il}D_\beta  \\
-\frac{i}{2\sqrt{2}} g b_c\delta_{jk}D_\alpha& \delta_{jl}\Pi_{\alpha\beta}\end{matrix} \right)\left(\begin{matrix}\Phi^k  \\
\Gamma^\beta_{l1}\end{matrix} \right)
\ee

If we call $\mathcal{M}$ this quadratic operator we have (see Appendix A)
\be
\det{\mathcal{M}}=\det(\delta_{ik}\mathcal{O})\det(\delta_{jl}\Pi_{\alpha\beta})\det(C_\alpha^{\,\,\beta}-\Pi^{-1
\gamma}_\alpha D_\gamma\mathcal{O}^{-1}D^\beta).
\ee

The determinant of $\Pi_{\alpha \beta}$  is given by
\begin{eqnarray}
\det \Pi_{\alpha \beta}&=&\det [\frac{1}{2}C^{\alpha\beta} C^ {\gamma \delta}\Pi_{\gamma \alpha} \Pi_{\delta\beta}]= 
\det [ \Theta_{\gamma \alpha} \Theta^{\gamma \alpha} +(g b_c)^2 \Theta_{\alpha}^{\,\,\alpha}+\frac{1}{2} (g b_c)^4]\nonumber \\
&=& \det [ \Box+ \frac{1}{8} (\alpha-1) (g b_c)^2D^2 -\frac{1}{8}\alpha (g b_c)^4]. \nonumber \\
\end{eqnarray}

For simplicity we work in the Landau gauge, $\alpha \rightarrow 0$. In this gauge, unless for multiplicative irrelevant constants,  we have
\be
\det \Pi_{\alpha\beta}= \det[D^2-\frac{1}{8}(g b_c)^2]\\.
\ee

As we can see from the result for $\ln\det \mathcal{O}$, below (by doing the substitution $\sigma_c \rightarrow (gb_c)^2$), this contribution starts at order $ (g b_c)^4$ and is so, out of the approximation that we are considering.

Taking into account the expression for the propagators of the superfields (\ref{prophi})-(\ref{propgamma0}) it can be shown that the last term in the expression of $\det \mathcal M$ has the form
\be
\det(C_\alpha^{\,\,\beta}-\Pi^{-1\gamma}_\alpha D_\gamma \mathcal{O}^{-1}D^{\beta})=
1-\alpha \frac{g^2b_c^2}{8} tr  \left((D^2-\alpha \frac{g^2b_c^2}{8})^{-1}(D^2-\sigma_C)^{-1}\right)+\mathcal O ((gb_c)^4)) . \\
\ee

In the Landau gauge, $\alpha \rightarrow 0$, the contribution to 
$\ln \det \Pi_{\alpha \beta}$ is zero up to the order $g^2$. So, the only contribution of $\det\mathcal M$ to the effective action reduces to
\bea
S_{1lp} ^{\Phi,\Gamma}&=&i (N-1)\log\det (D^2+\sigma_c)=i (N-1)\,tr \log (D^2+\sigma_c)\\ 
&=& \int d^2 \theta d^3x \frac {d^3p}{(2 \pi)^3} <\theta|<x||p><p|\ln(D^2+\sigma)|x>|\theta>
\eea
which, in the (K\"ahlerian) approximation $D_{\alpha} \sigma_c=\sigma_c D_{\alpha}$, results in:
%\be
%V_{1loop}\equiv- S_{1lp}/L^3 =N \left[\int d^2\theta \, \frac{\sigma^2_c}{8 \pi}+\mathcal{O}(g^4)+\mathcal{O}(1/N)\right]=N\frac{\sigma_1\sigma_2}{4 \pi}+\cdots
%\label{V1}
%\ee
\be
S_{1lp} =-\frac{NL^3}{8\pi}\int d^2\theta \, \sigma_c (\sigma_c^2)^
{1/2}==-L^3 N\frac{|\sigma_1|\sigma_2}{4 \pi}+\mathcal{O}(g^4) 
+\mathcal{O}(N^0)
\label{V1}
\ee

This simple form for the one-loop large N potential occurs also in nonsupersymmetric gauge theories (see \cite{Kang} for example).  
It can also be seen that this result is the first (linear) term in the expansion, of the exact one-loop calculation \cite{Maluf}
in powers of $\sigma_2$.

The last quadratic operator involving the remaining fields is given by
\be
\frac{1}{2}\left( \begin{matrix}  \Sigma & \varphi & \Gamma^\alpha_{11} \end{matrix} \right)\left( \begin{matrix} -1 & \sqrt{2 \lambda}b_c &0 \\
\sqrt{2 \lambda}b_c & 2 \mathcal{O} &-i\sqrt{2}b_c D_\beta\\
0&-i\sqrt{2}b_c D_\beta  &\tilde \Pi_{\alpha\beta} \end{matrix} \right)\left( \begin{matrix}\Sigma  \\
\varphi^\dagger\\
\Gamma^\beta_{11}  \end{matrix} \right)
\ee
but as can be seen, its contribution will be of subleading order in the $1/N$ expansion.

%%%%%%%%%%%%%%%%%%
%%%%%%%%%%%%%%%%%%
%%%%%%%%%%%%%%%%%%

\section{Two-loop contributions to the K\"ahlerian effective potential}

Following standard procedures and identities for the inversion of block matrix operators (see Appendix A), we can compute the superfield propagators:
\bea
\langle T \Phi_i (k,\theta)\Phi^\dagger_j (-k,\theta')    \rangle&=&-i\delta_{ij}\frac{D^2-\sigma_c}{k^2+\sigma_c^2}\delta^{(2)}(\theta-\theta')  +  \mathcal{O}(\alpha (g b_c)^2)\delta^{(2)}(\theta-\theta')  \label{prophi}\\
\langle T \Gamma^\alpha_{ij}(k,\theta) \Gamma^\beta_{kl}(-k,\theta')     \rangle&=&- \frac{i}{4}\delta_{il}  \delta_{kj}\left[\frac{D_\beta D_\alpha+\alpha D_\alpha D_\beta}{k^2}\right] \delta^{(2)}(\theta-\theta')\label{propgamma0}    \\
\langle T \Gamma^{*\alpha}_{1i}(k,\theta) \Gamma^\beta_{1j}(-k,\theta')     \rangle&=&- \frac{i}{4}\delta_{ij} 
\left[\frac{D_\beta D_\alpha+\alpha D_\alpha D_\beta}{k^2}\right] \delta^{(2)}(\theta-\theta')\label{propgamma0}.\\                                    \nonumber
\eea
%\langle T \Gamma^{*\alpha}_{1i}(k,\theta) \Gamma^\beta_{1j}(-k,\theta')     \rangle&=&- \frac{i}{2}\left[ \delta_{ij}  %\frac{(D^2-g^2b_c^2)D^2D_\beta D_\alpha}{k^2(k^2+g^4b_c^4)}-\right.\label{propgamma}\\
%&-&\left. \alpha\frac{ (D^2-\alpha g^2b_c^2)D^2D_\alpha D_\beta}{k^2(k^2+\alpha^2g^4b_c^4)} \delta^{(2)}(\theta-\theta') %+ \mathcal{O}(g^2)\right] \delta^{(2)}(\theta-\theta') \nonumber

In order to classify the possible vacuum diagrams contributing to the effective potential we must take into account the following observations:
 
Observation 1: Beyond two loops all diagrams contribute at order at least $g^3$; so, up to order $g^2$, it is enough to analyze two-loop diagrams.

Observation 2: Nonplanar diagrams are suppressed by factors of $1/N^2$ \cite{Hooft,Witten} and therefore it is only necessary to analyze the planar ones.

% since the highest possible order in the $1/N$ expansion is $N^2$.

\begin{figure}[h]
    \centering
    \includegraphics[width=0.7\textwidth]{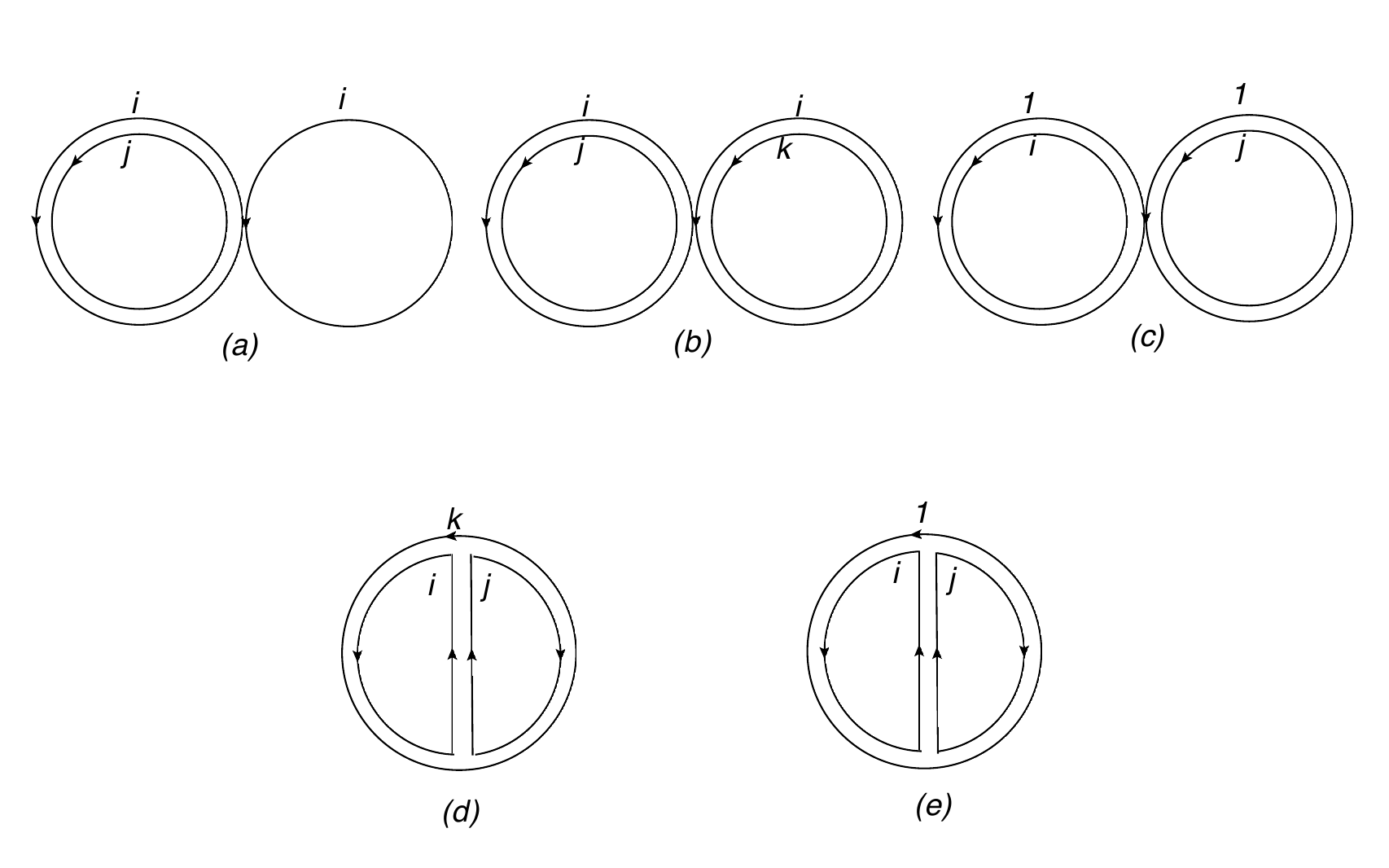}
    \caption{ Two-loop diagrams. (a), (b) and (d) are $\mathcal{O}(g^2)$ diagrams. (c) and (e) are $\mathcal{O}(g^4)$ diagrams.}
    \label{fig1}
\end{figure}

\begin{figure}[h]
    \centering
    \includegraphics[width=0.4\textwidth]{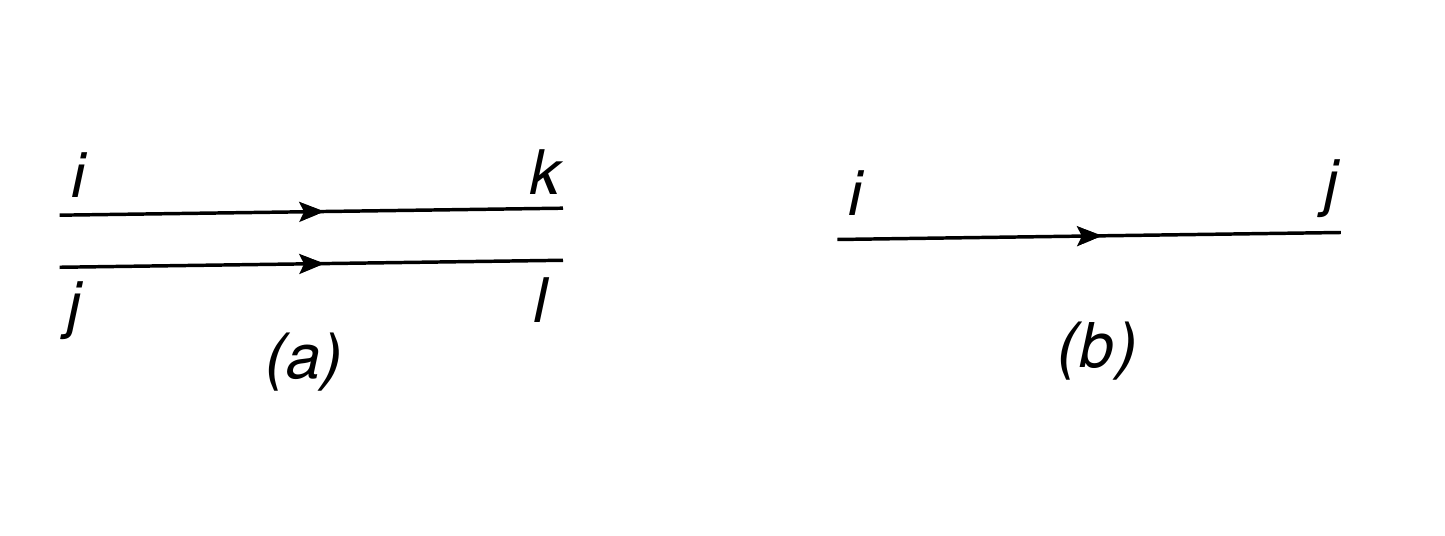}
    \caption{(a) represents the superfield propagator $\langle \Gamma^\alpha_{ij}(k,.\theta) \Gamma^\beta_{kl}(-k,\theta')\rangle$ while (b) represents the propagator $\langle \Phi_i (k,\theta)\Phi^\dagger_j(-k,\theta)  \rangle$.}
    \label{fig2}
\end{figure}

Let us start by analyzing the set of diagrams in Fig. \ref{fig1}, where $i,j,k,l=2,\cdots,N$, and the double line notation of  't Hooft \cite{Hooft} for the gauge fields is used (see Fig. \ref{fig2}).

Using the expression for the superfield propagators (\ref{prophi})-(\ref{propgamma0}) we get the following results:\\
Figure 1(a): $\mathcal{O}(g^2)$ and $\mathcal{O}(N)$, vanishing contribution in dimensional regularization. \\
Figure 1(b): $\mathcal{O}(g^2)$ and $\mathcal{O}(N^2)$, vanishing contribution in dimensional regularization.\\
Figure 1(c): $\mathcal{O}(g^4)$ and $\mathcal{O}(N)$, we will disregard it.\\
Figure 1(d): $\mathcal{O}(g^2)$ and $\mathcal{O}(N)$, vanishing contribution in dimensional regularization.\\
Figure 1(e) $\mathcal{O}(g^4)$ and $\mathcal{O}(N)$, we will disregard it.\\

\begin{figure}[h]
    \centering
    \includegraphics[width=0.4\textwidth]{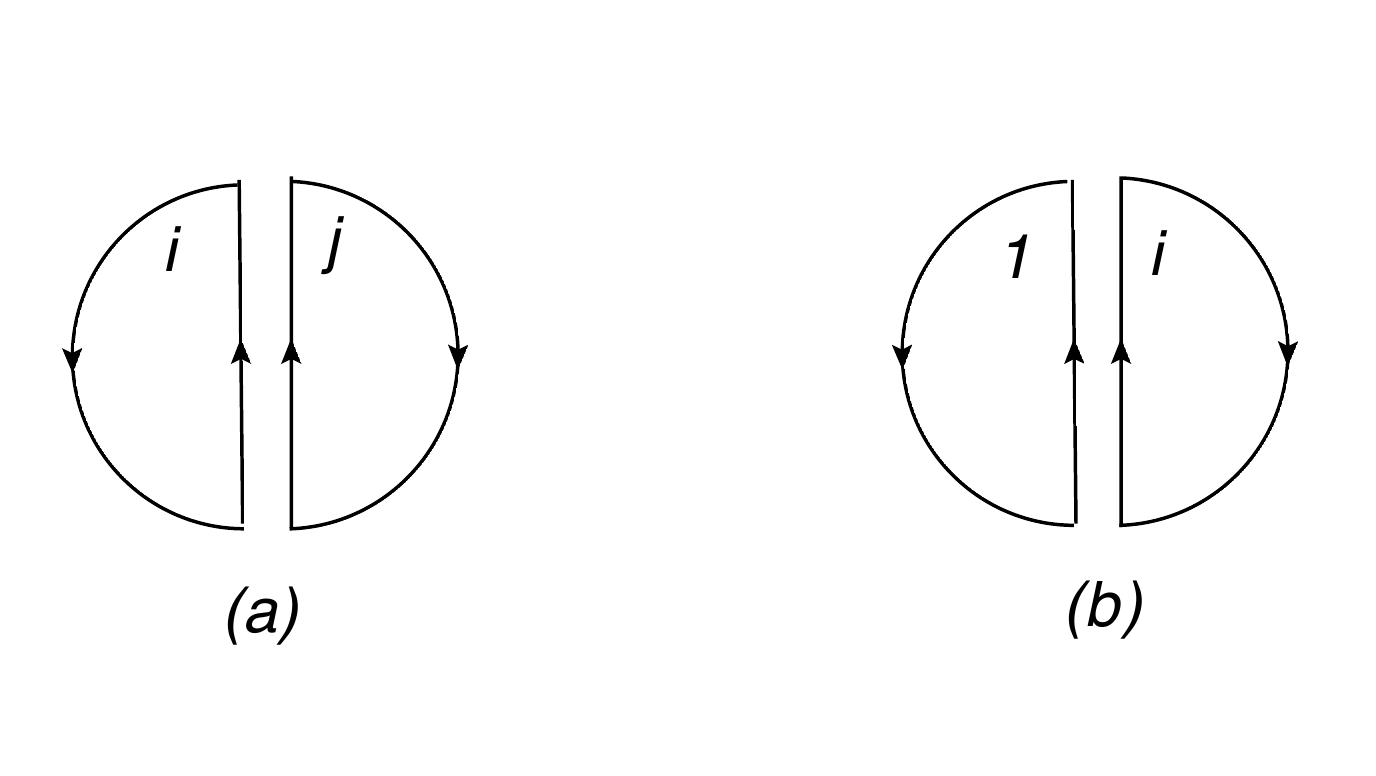}
    \caption{Nonvanishing two-loop diagram (a) and nonvanishing subleading two-loop diagram (b).}
    \label{fig3}
\end{figure}

The remaining diagrams to analyze are depicted in Fig. \ref{fig3}. Figure \ref{fig3}(b) is of order $N^0$; so, the only nonvanishing contribution to the effective potential is given by Figure \ref{fig3}(a),  (for details see Appendix B),

\be
D_{3(a)}=\frac{g^2}{8}N\int d^2\theta \frac{d^3 k d^3 p}{(2 \pi)^{6}}\frac{\sigma^2_c + kp}{(k^2+\sigma^2_c)(p^2+\sigma^2_c)(p+k)^2}+\mathcal{O}(N^0)
\ee

The integral in the momenta is divergent. By using dimensional reduction  \cite{Semenoff}, through the substitution $d=3-\epsilon$, and  introducing an arbitrary mass scale, by the substitution $d^3 k/(2 \pi)^3 \rightarrow \mu^{\epsilon} d^dk/(2 \pi)^d$, we obtain
\be
L^{-3} N^{-1}S_{2lp}=\int d^2\theta \frac{g^2}{64\pi^2}\frac{\sigma_c^2}{2} \left\{\frac{1}{3-d}+2-\gamma_E-\log\frac{\sigma_c^2}{4\pi\mu^2}\right\}\label{V2}.
\ee

Now by adding (\ref{V0}), (\ref{V1}) and (\ref{V2}) we have
\bea
L^{-3} N^{-1}S&=&\int d^2\theta \left[-\frac{1}{4}D^\alpha b_c D_\alpha b_c+\frac{\sigma_c b_c^2}{2}-
\frac{\sigma_c (\sigma_c^2)^{1/2}}{8 \pi}\right] \\
&+&\int d^2\theta \frac{\sigma_c^2}{2}\left[ -\frac{1}{\lambda_0}
+\frac{ g_0^2}{64 \pi^2}\left( \frac{1}{3-d} -\gamma_E+2-\log\frac{\sigma_c^2}{4\pi\mu^2}\right)\right]\\
\nonumber
\eea
where we called $\lambda_0$ and $g_0$ the unrenormalized coupling constants.

%%%%%%%%%%%%%%
%%%%%%%%%%%%%%
%%%%%%%%%%%%%%

\section{Renormalization}

First of all, we can rewrite the effective potential in component fields by integrating over $d^2\theta$ (taking the Grassmannian measure normalized as $\int d^2\theta\,\theta^2=-1$). The connection between the effective action and the K\"ahlerian effective potential is given by $S=-\int d^5zK_{eff}$; the effective potential is, by definition $V_{eff}=\int d^2 \theta K_{eff}$. After integration in $\theta$ we get
\bea
N^{-1}V_{eff}(\sigma_1,\sigma_2,b_1,b_2)&=&-\frac{1}{2}b_2^2- \frac{1}{2}(\sigma_2 b_1^2+2b_1b_2\sigma_1)
+\frac{1}{\lambda_0}\sigma_1\sigma_2+\frac{|\sigma_1|\sigma_2}{4 \pi}-\\
&-&g_0^2 \frac{\sigma_1\sigma_2}{(8\pi)^2}\left(\frac{1}{3-d}-\gamma_E+1-\log\frac{\sigma_1^2}{4\pi\mu^2}\right)+\mathcal{O}(g^4)+\mathcal{O}(1/N).\nonumber
\eea

Before solving the gap equations, we need to renormalize the effective potential. Up to $\mathcal{O}(g^2)$, the gauge coupling $g_0$ does not need an infinite renormalization and for simplicity we choose the finite renormalized gauge coupling as $g=g_0$; 
the finite renormalized $\lambda$ is chosen as
\be
\frac{1}{\lambda}=\frac{1}{\lambda_0}-\frac{g^2}{64\pi^2}\left[\frac{1}{3-d}+1-\gamma_E+\log(4\pi)\right].
\ee

With these choices the renormalized effective potential has the following form
\bea
N^{-1}V_{eff}^R(\sigma_1,\sigma_2,b_1,b_2)&=& -\frac{1}{2}b_2^2 -\frac{1}{2}\sigma_2 b_1^2-b_1b_2\sigma_1
+\frac{|\sigma_1|\sigma_2}{4\pi}\nonumber\\
&+&\frac{\sigma_1\sigma_2}{\lambda} +
\frac{g^2}{64\pi^2}\, \sigma_1 \sigma_2 \log[\frac{\sigma_1^2}{\mu^2}]+ 
\mathcal{O}(1/N)+\mathcal{O}(g^4).\label{potef}
\eea

As the gauge coupling constant $g$ (related to the CS level parameter by the substitution $\Gamma^{\alpha} \rightarrow g^{-1/2}\Gamma^{\alpha}$) does not need an infinity renormalization; it does not run with the energy scale, in agreement with other authors' results \cite{Semenoff}. For the matter superfield self-coupling, $\lambda$, we can see from the expression (\ref{potef}) that
\be
\left[\mu^2\frac{\partial}{\partial \mu^2} + \left(\frac{g}{8\pi}\right)^2 \lambda^2 \frac{\partial}{\partial \lambda}\right]V_{eff}=0,
\ee
from which it immediately follows that
\be
\lambda(\mu')=\frac{\lambda(\mu)}{1- \lambda(\mu)(\frac{g}{8\pi})^2\log\frac{\mu'^2}{\mu^2}}.
\ee
%We must observe that the singularity in $\lambda\rightarrow0$ is only aparent. For $\lambda=0$ we do not need to introduce the auxilary field $\sigma_c$, by adding the term in eq. (2.9). For $\lambda=0$ we must also consider $\sigma_c=0$
%and the above expression reduces to the trivial zero loop result.

This result shows that in the absence of the interaction with the CS field \lbrack that is, in the pure $U(N)$ vector matter model\rbrack, $\lambda$ does not need an infinite renormalization and does not run, in agreement with previous authors' results  \cite{BardeenS}, in the presence of the interaction with the CS; instead, it runs with the result $\lambda \rightarrow 0$ for $\mu' \rightarrow 0$ and has a Landau pole for $\mu'/\mu$ big enough . If this result remains in the exact leading $1/N$ approximation, or is an artifact of the truncation of the series in $\mathcal{O}(g^2)$, can only be decided by a higher order calculation. (work in progress) \cite{QS}.

An observation is important. The singularity in (\ref{potef}), when $\lambda \rightarrow 0$, is an artifact of the way we defined the auxiliary field $\Sigma$ (and their expectation value $\sigma_c$). Our choice was convenient to simplify the renormalization of the coupling constants, not requiring a wave function renormalization. For $\lambda=0$ the introduction of $\Sigma$ and $\sigma_c$ through the addition of the term (\ref{saux}) would not even be needed at all. To make this fact more explicitly, from now on we will redefine the field
 $\sigma_c$ as  $\sigma_c \rightarrow \lambda\sigma_c$ and make an additional finite renormalization of $1/\lambda$ to absorb an extra $(g^2/64\pi^2)\ln(\lambda^2)$ factor. In terms of these new fields the effective potential becomes
\bea
N^{-1}V_{eff}^R(\sigma_1,\sigma_2,b_1,b_2)&=& \lambda \sigma_2 \left[- \frac{b_1^2}{2} +\sigma_1+
\hat \lambda |\sigma_1|+{\hat g}^2 \lambda \sigma_1 \ln \frac{\sigma_1^2}{\mu^2}\right]\nonumber\\
&&-\frac{b_2^2}{2}-\lambda \sigma_1 b_1 b_2 +\mathcal{O}(1/N)+\mathcal{O}(g^4),\label{potef2}
\eea 
where we defined $\hat \lambda\equiv|\lambda|/4\pi>0$ and $\hat g\equiv g/8\pi$.

\noindent

\subsection{Gap equations and mass generation}
The gap equations corresponding to the above $V_{eff}^R$ are given by
\bea
0=\frac{\partial V}{ \partial \sigma_2}&=& -\lambda \frac{b_1^2}{2}+ \lambda \sigma_1  \left[1+\hat \lambda \epsilon( \sigma_1)+\lambda {\hat g}^2 \ln \frac{\sigma_1^2}{\mu^2}\right] \label{gaps2}\\
0=\frac{\partial V}{ \partial \sigma_1}&=&- \lambda b_1b_2+\lambda \sigma_2  \left[1+ \hat \lambda \epsilon(\sigma_1)                  +\lambda {\hat g}^2 \ln \frac{\sigma_1^2}{\mu^2}\right] \\
0=\frac{\partial V}{ \partial b_2}&=&-b_2-\lambda b_1\sigma_1\label{gapb2}\\
0=\frac{\partial V}{\partial b_1}&=& -\lambda \sigma_1b_2-\lambda \sigma_2 b_1\label{gapb1}
\eea
where $\epsilon(\sigma_1)$ is the sign of $\sigma_1$. 

Our expression $(\ref{potef2})$ for the effective potential  has the form of a perturbative (in the coupling $\hat g$) correction to the leading $1/N$ potential, of the pure SUSY $U(N)$ vector matter model. As discussed in \cite{Murphy}, the solutions of these gap equations  must be chosen as perturbative corrections (in ${\hat g}^2$) to that of the pure matter model;  
this model have been studied, in the last 30 years, by several authors, by using different methods, as for example, variational approximation in \cite{BardeenS}, $1/N$ approximation in \cite{Gudmund,Zinn} and functional renormalization group analyses in \cite{Wipf}. 
Using Eqs. (\ref{gapb2}) and (\ref{gapb1}) to eliminate $b_2$ and $\sigma_2$, we have the effective potential:
\be
V_{eff}=\frac{\lambda^2}{2} \sigma_1^2 b_1^2 \geqslant 0.\label{eepp}
\ee

In this expression, $\sigma_1$ and $b_1$ are related by (\ref{gapb1}). The SUSY preserving minima ($V_{eff}=0$) occur for the directions: $\sigma_1=0$ and $b_1=0$.

The possible phases that the model can have are

(a) For $\hat \lambda\equiv |\lambda|/4\pi \neq1$, starting with the line of minima $\sigma_1=0$ (or with $b_1=0$), as consequence of the gap equations, we have $\sigma_1=\sigma_2=b_1=b_2=0$. 
This solution corresponds to a phase in which SUSY and $U(N)$ symmetry are preserved.

(b) Besides the solution a), for $\hat \lambda= 1$, we can also have the solution $b_1=b_2=0$ and $\sigma_1=-\mu$, arbitrary. As a consequence of the K\"ahlerian  approximation, the value of $\sigma_2$ does not get determined by the gap equations, but  from the fact that the minimum of the potential is zero (which implies that SUSY is preserved), its value can be inferred to be $\sigma_2=0$.
In this phase, the mass of the fermionic component of the matter superfield $m_F^2=(\lambda \sigma_1)^2$ and the mass of the bosonic matter component, $m_B^2=\lambda^2 \sigma_1^2 -\lambda \sigma_2=m_F^2=\mu^2$ are equal and non-null. We have mass generation for the matter fields and breaking of the $U(N)$ symmetry and the scale symmetry. 

We must observe that for $\hat \lambda=1$, the solution $\sigma_1 \neq 0$, arbitrary, is already present in the pure matter model ($g=0$) \cite{BardeenS}. The new fact, introduced by the coupling to the CS field, is that this value is the scale parameter ($\mu$) introduced in the definition of  the dimensional reduction regularization. If we solve the gap equations (\ref{gaps2})-(\ref{gapb1}) for $g\neq 0$ we obtain again a massless and massive phase. In the massive phase only $\sigma_1$ is nonzero, and, since in our approximation $\sigma_1$ must lie around the mass scale $\mu$ and $\hat g$ small, the value of $\hat \lambda$ is constrained to be close to 1.
 
Bardeen \textit{et al} \cite{BardeenS} studied the pure matter model using a variational method and ultraviolet cutoff regularization. Their model includes also a mass term in the classical action, whose mass coefficient they call $\mu$. Our model corresponds to their particular case $\mu=0$, in which the model is classically scale invariant. Our two phases are in agreement with their results (for the pure matter model), for this choice of their parameter $\mu$.

\section{The dilatino pole}

As we found in the previous section, for $g= 0$ and $\lambda=4\pi$ the scale invariance is spontaneously broken. This broken phase also appears for $g\neq 0$ and $\lambda\sim4\pi$. The condition that $\lambda$ must lie around this critical value arises from the fact that our calculation is valid at $g^2$ order. Of course,  one should expect that in the exact calculation in $g$, the critical value of $\lambda$ is not necessarily close to $4\pi$, provided that $g$ is sufficiently large. We have also found that the effective potential has a vanishing value at its minimum and therefore, the ground state of the model is supersymmetric. This implies that the Goldstone boson associated with this breaking (the dilaton) must be accompanied by its supersymmetric partner, a Goldstone fermion called a dilatino \cite{BardeenS}. The dilatino pole must occur at $p^2=0$ and can be found in the fermion-boson scattering amplitude. Before calculating this amplitude we need the explicit form of the action in components. The relevant part of it, once we eliminated the auxiliary fields, can be written as follows:
\be
S=\int d^3x \left[2 \epsilon^{\mu\nu\rho}\text{Tr}\left(A_\mu\pa_\nu A_\rho-\frac{2i }{3}gA_\mu A_\nu A_\rho\right)+D^\mu\bar{\phi}D_\mu\phi-\bar{\psi}\gamma^\mu D_\mu\psi-\lambda \left(\bar{\psi}\phi\right) \left(\bar{\phi}\psi\right)+...\right]
\ee
where the dots stand for terms which do not contribute at leading order to the fermion-boson scattering.  The amplitude can be written as follows (Fig. \ref{dila}):
\begin{figure}[h]
    \centering
    \includegraphics[width=0.6\textwidth]{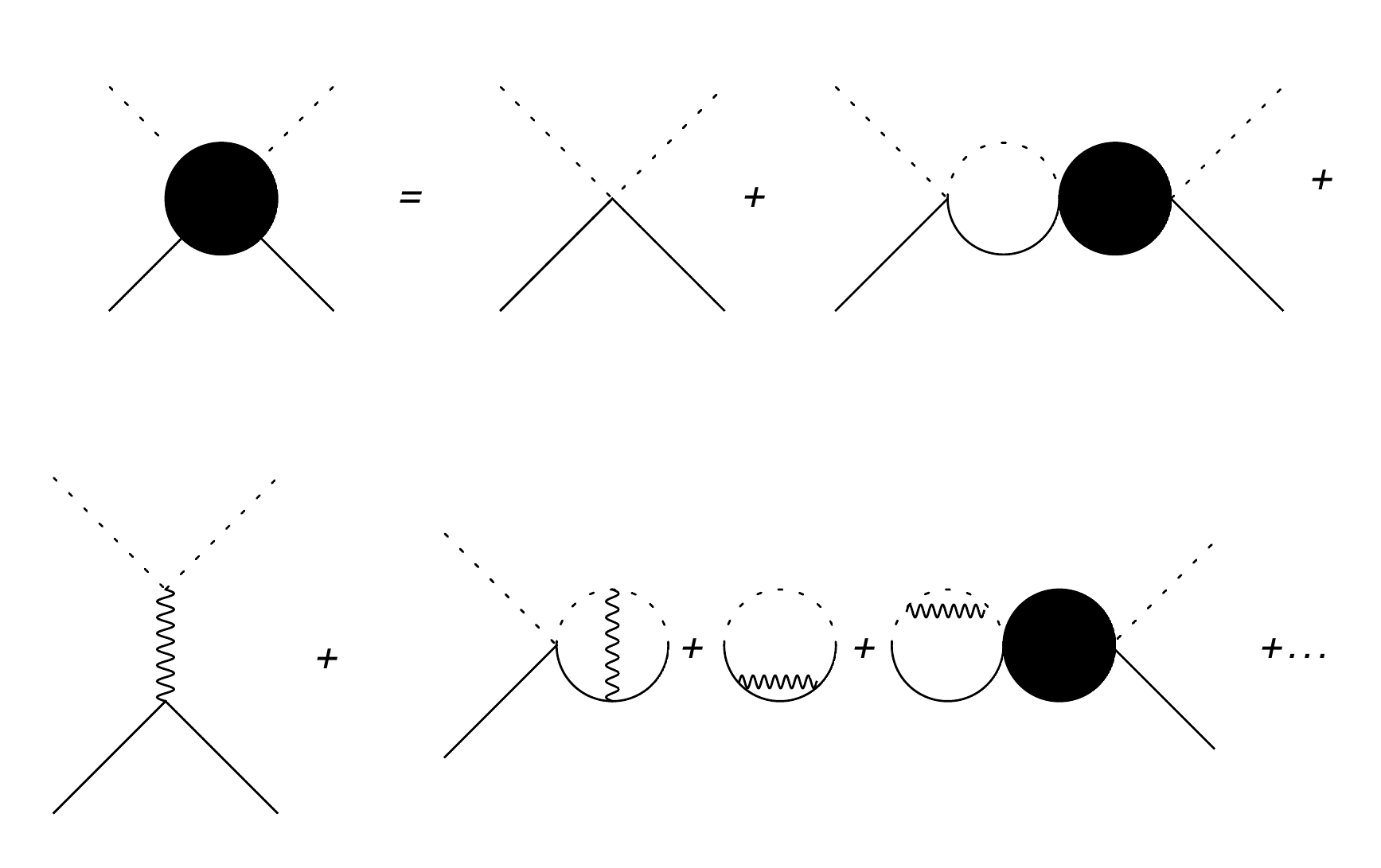}
    \caption{ Fermion-boson scattering amplitude at leading order. Dashed line fermion; solid line, boson}
    \label{dila}
\end{figure}

\bea
\Gamma\left[p^2\right]&=&\frac{2\lambda}{N}+2\lambda\int \frac{d^3 k}{(2\pi)^3}\frac{\slashed{p}+\slashed{k}-\mu}{(p+k)^2+\mu^2}\left(\frac{1}{k^2+\mu^2}\right)\Gamma\left[p^2\right]+\frac{g^2}{N}f(p)+\\
&+& g^2\int   \frac{d^3 k}{(2\pi)^3} \frac{d^3 l}{(2\pi)^3}h(p,k,l)\Gamma\left[p^2\right]
\eea
where $\mu^2=\left(\lambda\sigma_1\right)^2$, and the functions $f(p)$ and $h(p,k,l)$ are associated with the last two diagrams in Fig. \ref{dila}.  If we expand for small $p$ and solve for $\Gamma\left[p^2\right]$ we obtain
\be
\Gamma\left[p^2\right]=\frac{1}{N}\left(2\lambda+g^2 f(p)\right)\left(1-\frac{\lambda}{4\pi}-g^2 A +\slashed{p}\left(\frac{1}{2\pi\sigma_1}-\frac{g^2 B}{\sigma_1}\right)\right)^{-1}\label{ampli}
\ee
where $A$ and $B$ are numerical constants coming from the expansion in $p$ of the function $h(p,k,l)$, but irrelevant for our purposes. For the case $g=0$, we see from the denominator of (\ref{ampli}), that the pole appears at $p^2=0$ for the critical value $\lambda=4\pi$ ($\hat{\lambda}=1$). This result is in complete agreement with the one obtained in \cite{BardeenS} for the supersymmetric $\varphi^6$ model. The new fact is that for $\lambda\neq 4\pi$, the pole is still located at $p^2=0$ while the constant $g$ takes the value
\be
g^2=\frac{1}{A}\left(\frac{\lambda}{4\pi}-1\right)
\ee
(since in our case $g$ is small, $\lambda$ must be close to the critical value). Now, since the ground state is supersymmetric, the pole of the dilaton must occur also at $p^2=0$. Therefore, even when the gauge coupling is nonzero, we find a dilaton and dilatino associated with the breaking of scale invariance.

\section{Summary}
In this work we presented a large N expansion of the non-Abelian SUSY Chern-Simons-matter model. We determined the superpropagators in the large N limit and order $g^2$ for all the fields. After the shifting of the superfields we have obtained the effective potential in the K\"ahler approximation. Such effective potential is exact at leading order in the $1/N$ expansion, for finite $\lambda$ (the marginal coupling constant) and at order $g^2$ in the gauge coupling. 

Once the effective potential was determined we solved the gap equations.
We showed that there exist two phases, a massless one, where the superconformal symmetry is preserved and a massive one. The massive phase is characterized by the marginal coupling constant $\lambda$, such that the complex field  $\phi_a$ and the fermion $\psi_\alpha^a$ become massive for arbitrary $\lambda$, and the gauge fields remain massless. In the limit $g\rightarrow 0$ the model becomes a free SUSY Chern-Simons term plus a SUSY $\varphi^6$ theory in three dimensions. In this limit we obtained again a massless phase for all values of the coupling constant $\lambda$. But, for the fixed value $\lambda=4\pi$ we found also a massive phase where the superconformal symmetry is broken. In this point the fields $\phi_a$ and $\psi_\alpha^a$ can acquire mass, but the gauge fields remain massless. After integrating over the Grassmann coordinates and eliminating the auxiliary fields the effective coupling of the $\varphi^6$ term is $\lambda_c^2=(4\pi)^2$ which coincides with the one obtained in \cite{Bardeen2,BardeenS} for the $ \varphi^6$ model with mass term and quartic interaction. 

The position of the minimum of the effective potential (\ref{eepp}) does not depend on the gauge coupling (or the Chern-Simons level $\kappa=4\pi/g^2$), but this must be an artifact of the perturbative expansion in $g$. An exact calculation in the gauge coupling must show this dependence even at leading order  in  the $1/N$ expansion.  

 If we extend the supersymmetry from $\mathcal{N}=1$ to $\mathcal{N}=2$ (which corresponds to the constraint $\lambda=g^2/4$, \cite{Ivanov,Weinberg}), the model possesses a massless phase, but the massive phase is out of the perturbative regime in the gauge coupling $g$. 
 We found also that associated with the breaking of scale invariance, a massless dilatino appears in the theory as a  $\phi\psi$ state, and due to the supersymmetric invariance of the ground state, we can ensure that the dilaton is also massless. This is in complete agreement with the results obtained in \cite{BardeenS} for the supersymmetric $\varphi^6$ model.
  
In conclusion we found that the dynamical breakdown of superconformal symmetry can occur in the $\mathcal{N}=1$ large $N$ limit of the Chern-Simons-matter  theory.
 
 The superfield formalism provides a nice framework for the study of effective potentials, adding both, bosonic and fermionic contributions in a single superfield. Further investigations
of the SCS model (e.g., $\mathcal{N}=2$ model, subleading correction in the large $N$ expansion to the effective potential, etc.) will
be pursued in future works.

{\bf Acknowledgements.} The work of J.M.Q. is supported by Funda\c{c}\~ao de Amparo \`{a} Pesquisa do Estado de S\~ao Paulo (FAPESP). The work of A.J.S is partially supported by Conselho Nacional de Desenvolvimento Cient\'{i}fico e Tecnol\'{o}gico (CNPq).

\section{Appendix A: Some useful Identities}

Let  $A,B,C,D,X,Y$ be operators and $\epsilon\in\mathbb{R}$.  Let $M$ be the following matrix operator 
\[ M=\left( \begin{array}{cc}
A & B  \\
C&D\\
\end{array} \right).
\]
\\
 We have
\bea
\det M=\det[A] \det[D-B A^{-1}C]&=&\det[A] \det[D] \det[I-D^{-1} B A^{-1} C]\\
\det[X+\epsilon Y]&=&\det[X](1-tr[X^{-1}Y]\epsilon)+\mathcal{O}(\epsilon^2)
\eea
and for the inverse matrix
\[ M^{-1}= \left( \begin{array}{ccc}
(A-BD^{-1}C)^{-1} &-(A-BD^{-1}C)^{-1} BD^{-1}  \\
-D^{-1}C (A-BD^{-1}C)^{-1} &D^{-1}+D^{-1} C(A-BD^{-1}C)^{-1}BD^{-1} \end{array}  \right)\]

\section{Appendix B: two-loop supergraph }

Let us call the contribution of the Fig. 3(a), $G_{3(a)}$. In terms of superfields and superderivatives we have
\be
G_{3(a)}=-\frac{g^2}{8} \frac{(N-1)^2}{N}V\int_{\theta_1}\int_{\theta_2}\int_{p}\int_{k}\frac{\Omega^{\alpha\beta}(\theta_1,\theta_2)\Sigma_{\alpha}(\theta_1)\Sigma_{\beta}(\theta_2)}{\Delta(p,k,\sigma_c)} \label{D3a}
\ee 
where $V$ stands for the volume of space-time and 
\bea
\Omega^{\alpha\beta}(\theta_1,\theta_2)&=&\langle \Gamma^\alpha (\theta_1)\Gamma^\beta(\theta_2)   \rangle\\
\Sigma_{\alpha}(\theta_1)&=&D_{1\alpha}\Phi^\dagger(\theta_1)\Phi(\theta_1)-\Phi^\dagger(\theta_1)D_{1\alpha}\Phi(\theta_1)\\
\Delta(p,k,\sigma_c) &=&(p^2+\sigma_c^2)(k^2+\sigma_c^2)(p+k)^2
\eea

The superfields $\Phi$ are understood to be the $\Phi^a,a=1,..,N$. The term $\Sigma_\alpha \Sigma_\beta$ is contracted into superpropagators
\bea
\Sigma_\alpha (\theta_1) \Sigma_\beta(\theta_2)&=&-\langle \Phi (\theta_1)D_\beta \Phi^\dagger(\theta_2)\rangle \langle \Phi (\theta_2)D_\alpha \Phi^\dagger(\theta_1)\rangle+\\
&+&\langle \Phi(\theta_1)\Phi^\dagger(\theta_2)   \rangle   \langle D_\beta \Phi(\theta_2)D_\alpha\Phi^\dagger(\theta_1)   \rangle-\nonumber\\
&-& \langle   \langle D_\beta \Phi(\theta_1)D_\alpha\Phi^\dagger(\theta_2)   \rangle\langle \Phi(\theta_2)\Phi^\dagger(\theta_1)   \rangle+\nonumber\\
&+& \langle D_\alpha \Phi(\theta_1)\Phi^\dagger(\theta_2)  \rangle \langle D_\beta \Phi(\theta_2)\Phi^\dagger(\theta_1)      \rangle\nonumber
\eea

After integration by parts in the integral (\ref{D3a}) we can isolate the last Dirac delta in the numerator, and integrate over $\theta_2$.  By using the following $D$-algebra identities, 
\bea
D_{1\alpha(p)} D_{1\beta(p)}&=&p_{\alpha\beta}-C_{\alpha\beta}D_1(p)^2\\
D_1^\beta(p)D_{1\alpha}(p)D_{1\beta}(p)&=&0\\
\{D_{1\alpha}(p),D^2_{1}(p)\}&=&0\\
\delta_{12}D^2_1 (p)\delta_{12}&=&1\\
\delta_{12}D_{1\alpha} (p)\delta_{12}&=&0
\eea

Finally we can rewrite the contribution of the diagram in terms of usual momenta, 
\be
G_{3(a)}=-\frac{1}{8}g^2 \frac{(N-1)^2}{N}V\int_{\theta_1}	\int_p\int_k \frac{pk+\sigma_c^2}{\Delta(p,k,\sigma_c) } 
\ee

Now using the regularized integrals $\lbrack d^3k/(2 \pi)^3 \rightarrow \mu^{\epsilon} d^dk/(2 \pi)^d\rbrack$:
\bea
I(m_1,m_2,m_3)=\mu^{2 \epsilon}\int \frac{d^d k d^d p}{(2\pi)^{2d}}\frac{1}{(k^2+m_1^2)(p^2+m_2^2)((p+k)^2+m_3^2)}=\nonumber \\
\frac{1}{32\pi^2}\left \{  \frac{1}{\epsilon}-\gamma_E+1-\log\left[\frac{(m_1+m_2+m_3)^2}{4\pi\mu^2}\right]\right \}\\
J(m_1,m_2,m_3)=- \mu^{2 \epsilon} \int \frac{d^d kd^d p}{(2\pi)^{2d}}\frac{k p}{(k^2+m_1^2)(p^2+m_2^2)((p+k)^2+m_3^2)}=\nonumber \\
\frac{1}{32\pi^2}\left( m_1 m_2-m_2 m_3-m_1m_3  \right)+\frac{1}{2}(m_1^2+m_2^2-m_3^2)I(m_1,m_2,m_3)
\eea

we arrive at the expression (\ref{V2}).

\end{document}